\documentclass[preprint,aps,prd,showpacs,nofootinbib,superscriptaddress,tightenlines]{revtex4}

\def\slash#1{#1 \hskip-0.45em /}
\def\Slash#1{#1 \hskip-0.59em /}

\newcommand{\nn}{\nonumber}

\begin{document}

\title{\mbox{}\\[10pt]
The leading twist light-cone distribution amplitudes for the S-wave and 
P-wave quarkonia and their applications in single quarkonium exclusive productions}

\author{Xiang-Peng Wang\footnote{E-mail: wangxiangpeng11@mails.ucas.ac.cn}}
\author{Deshan Yang\footnote{E-mail: yangds@ucas.ac.cn}}
\affiliation{School of Physics, University of
Chinese Academy of Sciences, Beijing 100049, China\vspace{0.2cm}}

\date{July 7, 2017}
\begin{abstract}
{In this paper, we calculate twist-2 light-cone distribution amplitudes (LCDAs) of the S-wave and P-wave quarkonia (namely $^1S_0$  state $\eta_{Q}$, $^3S_1$ state $J/\psi(\Upsilon)$, $^1P_1$  state $h_{Q}$ and $^3P_J$ states $\chi_{QJ}$ with $J=0,1,2$ and $Q=c,b$) to the next-to-leading order of the strong coupling $\alpha_s$ and leading order of the velocity expansion $v$ in the non-relativistic QCD (NRQCD).  We apply these LCDAs to some single  quarkonium exclusive  productions at large center-of-mass
energy, such as $\gamma^*\to \eta_Q\gamma, \chi_{QJ}\gamma~(J=0,1,2)$, $Z\to \eta_Q\gamma, \chi_{QJ}\gamma~(J=0,1,2), J/\psi (\Upsilon)\gamma, h_Q\gamma$ and $h\to J/\psi \gamma$, by adopting the collinear factorization. The asymptotic behaviors of those processes obtained in NRQCD factorization are reproduced.
}
\end{abstract}

\pacs{\it 12.38.-t, 12.38.Cy, 12.39.St, 14.40.Gx}

\maketitle

\section{Introduction}

One of main fields for precision examination of the perturbative Quantum Chromodynamics (QCD) is the study of the hard exclusive processes with the large momentum transfer involved. The collinear factorization has been a well-established calculation framework for more than  three decades\cite{Lepage:1980fj,Chernyak:1983ej}. In this framework, the amplitudes of many hard exclusive processes can be expressed as convolutions of the perturbatively calculable hard-kernels and the universal light-cone distribution amplitudes (LCDAs), in which the short-distance and long-distance contributions are clearly factorized. For instance, the electromagnetic form-factor of $\gamma^*\gamma\to \pi^0$ at large momentum transfer can be expressed as
\begin{eqnarray}
F(Q^2)= f_\pi \int_0^1 dx \, T_H(x;Q^2,\mu) \phi_\pi(x;\mu)+{\cal O}(\Lambda^2_{\rm QCD}/Q^2)\,,
\end{eqnarray}
where hard-kernel $T_H(x;Q^2,\mu)$ contains the short-distance dynamics, while the LCDA of pion $f_\pi\phi_\pi(x;\mu)$ is a purely non-perturbative object parametrizing the universal hadronization effects around the light-like distance. The LCDAs for light hadrons are not perturbatively calculable,  one has to extract their informations from the experiments, or calculate or constrain them by various non-perturbative methods, such as QCD sum rules, Lattice simulations. However, the dependence of  these LCDAs on the renormalization scale $\mu$ are perturbatively calculable. For instance, renormalization scale dependence of twist-2 LCDA of pion is governed by the celebrated Efremov-Radyushkin-Brodsky-Lepage (ERBL) equation \cite{Lepage:1979zb, Efremov:1979qk} 
\begin{eqnarray}
\frac{d}{d\ln\mu^2}f_\pi \phi_\pi(x;\mu)=\frac{\alpha_s}{2\pi}C_F\int_0^1 V_0(x,y)f_\pi\phi_\pi(y;\mu)\,,
\end{eqnarray}
where $V_0(x,y)$ is the so-called Brodsky-Lepage kernel.

For the quarkonium involved exclusive processes, if the momentum transfer square is much greater than the mass square of the quarkonium, the collinear factorization can be invoked as well \cite{Bodwin:2008nf,Bodwin:2010fi}.  Many phenomenological applications along this line have been made for exclusive hard production of charmonium \cite{Ma:2004qf,Bondar:2004sv,Braguta:2008tg,Bodwin:2006dm,Braguta:2009df,Braguta:2009xu,Braguta:2010mf}, exclusive charmonium production in $B$ meson  decays \cite{Cheng:2000kt,Cheng:2001ez,Song:2002gw,Song:2002mh},  {\it etc}.  
All of these applications require the understanding of the LCDAs for quarkonia.

Different from the LCDAs for the light mesons which rely completely on the dynamics in the non-perturbative regime of QCD, one believes that the LCDAs for quarkonia can be further factorized into the products of the perturbatively calculable parts and non-perturbative behavior of the wave-functions of quarkonia at origin, due to the nature of quarkonium as a non-relativistic bound state of heavy quark and anti-quark. The standard theoretical tool to deal with the heavy quark bound state system is the NRQCD factorization \cite{Bodwin:1994jh,Brambilla:2004jw}, in which all information of hadronization of quarkonium is encoded in the NRQCD matrix elements. Thus, there must be connections between the LCDAs of quarkonia and NRQCD matrix elements. For examples,  in \cite{Braguta:2006wr,Braguta:2007fh,Braguta:2008qe}, the authors try to constrain their models for the LCDAs of quarkonia by relating the moments of LCDAs with the local NRQCD matrix elements; in \cite{Ma:2006hc,Bell:2008er}, the authors calculated the leading twist LCDAs of the S-wave quarkonia within the  NRQCD framework, and express the LCDAs in form of the products of perturbatively calculable distribution parts and lowest order NRQCD matrix-elements. 
 
Especially, the attempts in \cite{Ma:2006hc,Bell:2008er} open a way to connect the predictions of hard quarkonium exclusive productions within the collinear factorization directly to those made within the NRQCD factorization (for examples, the many theoretical calculations based on NRQCD factorization \cite{Braaten:2002fi, Liu:2002wq, Hagiwara:2003cw, Zhang:2005cha, Gong:2007db, He:2007te, Bodwin:2007ga, Sang:2009jc,Li:2009ki,Wang:2011qg, Dong:2011fb, Dong:2012xx, Li:2013qp, Dong:2013qw}, triggered by the recent experimental measurements of charmonium exclusive productions at $B$-factories \cite{Abe:2002rb,Pakhlov:2004au,Abe:2004ww}). In particular, in \cite{Jia:2008ep,Jia:2010fw}, the authors have shown that  the collinear factorization indeed can reproduce the exact asymptotic behavior of NRQCD predictions at the leading logarithms (LL) and next-to-leading order (NLO) of the strong coupling $\alpha_s$, respectively,  for a certain class of the quarkonium exclusive productions, if one employs the leading twist LCDAs calculated in \cite{Bell:2008er}; and the ERBL equations can be used to resum the large logarithms appearing in the NRQCD factorization calculations for the exclusive quarkonium productions, while such resummation cannot be done within the NRQCD factorization. 

As a successive work of  \cite{Jia:2008ep,Jia:2010fw}, in this paper, we calculate ten leading twist LCDAs for the S-wave and P-wave quarkonia, namely $^1S_0$, $^3S_1$, $^1P_1$ and $^3P_J$ ($J=0,1,2$) states, to the NLO of $\alpha_s$ and leading order of non-relativistic expansion parameter $v$,  by adopting methods developed in \cite{Ma:2006hc,Bell:2008er}.  For three LCDAs of S-wave quarkonia, we get slightly different  results from those obtained in \cite{Ma:2006hc}, and confirm the results of LCDA for $^1S_0$ state given in \cite{Bell:2008er} . The seven leading twist LCDAs of P-wave quarkonia at NLO are totally new. All of these leading twist LCDAs at NLO do obey the ERBL equations, and can be applied to various quarkonium involved hard exclusive processes.

This paper is organized as follows: in Sect.\ref{sect:definitions}, we give the definitions of the leading twist LCDAs for the S-wave and P-wave quarkonia, in terms of the matrix-elements of a certain class of non-local QCD operators, and their tree-level forms at the leading order of $v$; in Sect. \ref{sect:NLOLCDAs}, we present our main results of this paper, the LCDAs at the NLO of $\alpha_s$ and leading order of $v$; in Sect.\ref{sect:applications}, as applications and non-trivial examinations of our results, we calculate the $\gamma^*\to \eta_Q\gamma, \chi_{QJ}\gamma$, $Z\to \eta_Q\gamma, \chi_{QJ}\gamma\,(J=0,1,2)\,, J/\psi(\Upsilon)\gamma, h_Q\gamma$ and  $h\to J/\psi\gamma$ within the collinear factorization ,  by using the LCDAs we calculate, and show how we can reproduce  the asymptotic behavior of the NLO NRQCD predictions for those processes exactly; finally, we summarize our work in Sect.\ref{sect:summary}.

\section{The definitions of LCDAs for quarkonia\label{sect:definitions}}
\subsection{Notations}  

We adopt the following notations for the decompositions of momenta: the momentum of quarkonium $H$ is $P^\mu \equiv m_H v^\mu$ with $v^2=1$, and a 4-vector $a^\mu$ can be decomposed as $a^\mu = v\cdot a v^\mu+a_\top^\mu$ where $v\cdot a_\top\equiv 0$. We also use the same notation $v$ for the non-relativistic expansion parameter, which is typical size of the relative velocity of quark and anti-quark inside a quarkonium. In the context, one should not confuse these two.  We also introduce two light-like vectors $n_\pm^\mu$ such that $n_\pm^2=0$ and $n_+n_-=2$, and any 4-vector $a^\mu$ can be decomposed as $a^\mu=n_+a n_-^\mu/2+n_-a n_+^\mu/2+a_\perp^\mu$ with $n_\pm a_\perp\equiv 0$. For convenience, we set $v^\mu=(n_+v n_-^\mu+n_-v n_+^\mu)/2$ (apparently $n_+v n_-v =1$). 

\subsection{Defintions of the LCDAs}
The leading twist, i.e. twist-2,  LCDAs for the S-wave and P-wave quarkonia are
defined as the matrix-elements of the proper gauge-invariant non-local quark bilinear operators
\begin{eqnarray}\label{eq:string}
J[\Gamma](\omega)\equiv(\bar Q W_c)(\omega n_+/2)\slash n_+\Gamma
(W_c^\dagger Q)(-\omega n_+/2)\,,
\end{eqnarray}
where $Q$ is the heavy quark field in QCD, the Wilson-line 
\begin{eqnarray}
W_c(x)&=&{\rm P}\exp\left(i g_s\int_{-\infty}^0 ds
n_+A(x+sn_+)\right)\,,
\end{eqnarray}
is a path-ordered exponential with the path along the $n_+$ direction, $g_s$ is the SU(3) gauge coupling and $ A_\mu (x)\equiv A^a_\mu(x) T^a$ ($T^a$ are the generators of SU(3) group in the fundamental representation). 

The ten non-vanishing twist-2 LCDAs of the S-wave and P-wave quarkonia  are defined as\footnote{Here we follow the definitions of the LCDAs for P-wave mesons in series papers by K.C. Yang {\it et al} \cite{Yang:2005gk,Cheng:2005nb,Yang:2007zt,Cheng:2010hn},  by setting $z=\omega n_+/2$, and $p^\mu=n_+Pn_-^\mu/2$. Thus $p\cdot z\equiv n_+P \omega/2$.}
\begin{eqnarray}
&&\langle H(^1S_0,P)\vert J[\gamma_5](\omega)\vert 0\rangle=- i f_{P} n_+P\int_0^1dx\,e^{i\omega n_+P(x-1/2)}
\hat{\phi}_{P}(x;m,\mu)\,,\\
&&\langle H(^3S_1,P,\varepsilon^*)\vert J[1]\vert 0(\omega)\rangle=-i f_{V} m_{V}n_+\varepsilon^*\int_0^1dx\,e^{i\omega n_+P(x-1/2)}
\hat{\phi}_{V}^{\parallel}(x;m,\mu)\,,\\
&&\langle H(^3S_1,P,\varepsilon^*)\vert J[\gamma_\perp^\alpha](\omega)\vert 0\rangle=-i f_{V}^{\perp}n_+P\varepsilon^{*\alpha}_\perp\int_0^1dx\,e^{i\omega n_+P(x-1/2)}
\hat{\phi}_{V}^{\perp}(x;m,\mu)\,,\\
&&\langle H(^1P_1,P,\varepsilon^*)\vert J[\gamma_5](\omega)\vert 0\rangle=i f_{1A} m_{1A}n_+\varepsilon^{*} \int_0^1dx\,e^{i\omega n_+P(x-1/2)}
\hat{\phi}_{1A}^{\parallel}(x;m,\mu)\,,\\
&&\langle H(^1P_1,P,\varepsilon^*)\vert J[\gamma_\perp^\alpha\gamma_5](\omega)\vert 0\rangle= i f_{1A}^{\perp}n_+P\varepsilon^{*\alpha}_\perp\int_0^1dx\,e^{i\omega n_+P(x-1/2)}
\hat{\phi}_{1A}^{\perp}(x;m,\mu)\,,\\
&&\langle H(^3P_0,P)\vert J[1](\omega)\vert 0\rangle= f_{S} n_+P\int_0^1dx\,e^{i\omega n_+P(x-1/2)}
\hat{\phi}_{S}(x;m,\mu)\,,\\
&&\langle H(^3P_1,P,\varepsilon^*)\vert J[\gamma_5](\omega)\vert 0\rangle=i f_{3A} m_{3A}n_+\varepsilon^{*}\int_0^1dx\,e^{i\omega n_+P(x-1/2)}
\hat{\phi}_{3A}^{\parallel}(x;m,\mu)\,,\\
&&\langle H(^3P_1,P,\varepsilon^*)\vert J[\gamma_\perp^\alpha\gamma_5](\omega)\vert 0\rangle=i f_{3A}^{\perp} n_+P\varepsilon^{*\alpha}_\perp\int_0^1dx\,e^{i\omega n_+P(x-1/2)}
\hat{\phi}_{3A}^{\perp}(x;m,\mu)\,,\\
&&\langle H(^3P_2,P,\varepsilon^*)\vert J[1](\omega)\vert 0\rangle= f_{T} \frac{m_T^2}{n_+P} n_{+\alpha}n_{+\beta}\varepsilon^{*\alpha\beta} \int_0^1dx\,e^{i\omega n_+P(x-1/2)}
\hat{\phi}_{T}^{\parallel}(x;m,\mu)\,,\\
&&\langle H(^3P_2,P,\varepsilon^*)\vert J[\gamma_\perp^\alpha](\omega)\vert 0\rangle=f_{T}^{\perp} m_T n_{+\rho}\varepsilon^{*\rho\alpha_\perp}\int_0^1dx\,e^{i\omega n_+P(x-1/2)}
\hat{\phi}_{T}^{\perp}(x;m,\mu)\,,
\end{eqnarray}
where $f$, $\varepsilon^*$ and $\hat \phi(x)$ are decay constants, polarization vectors/tensors, and twist-2 LCDAs of corresponding quarkonia, respectively. $x$ denotes the light-cone fraction, and $\mu$ is the renormalization scale. In whole of this paper, we will also adopt the notation $\bar x\equiv 1-x$ for any light-cone fraction $x\in [0,1]$.

Due to the discrete $C,\,P,$ and $T$ symmetries, one can check that, when $\omega\to 0$, we have
\begin{eqnarray}
&&\int_0^1 dx \hat\phi_{1A,T}^\parallel (x)=\int_0^1 dx \hat\phi_{3A,T}^\perp(x)=\int_0^1 dx \hat\phi_{S}(x)=0\,,
\end{eqnarray}
and corresponding integrals of the rest LCDAs do not vanish. Thus, 
 we set the normalization conditions for the LCDAs as following
\begin{eqnarray}\label{eq:norm1}
&&\int_0^1 dx \hat\phi_{P} (x)=\int_0^1 dx \hat\phi_{V}^\parallel (x)=\int_0^1 dx \hat\phi_{V}^\perp(x)=\int_0^1 dx \hat\phi_{1A}^\perp (x)=\int_0^1 dx \hat\phi_{3A}^\parallel (x)=1\,,
\end{eqnarray}
\begin{eqnarray}
\label{eq:norm2}
&&\int_0^1 dx \hat\phi_{1A,T}^\parallel (x)(2x-1)=\int_0^1 dx \hat\phi_{3A,T}^\perp(x)(2x-1)=\int_0^1 dx\hat\phi_{S}(x)(2x-1)=1\,.
\end{eqnarray}
Then, some decay constants defined above can be related to the following matrix-elements of local operators
\begin{eqnarray}
&&\langle H(^1S_0,P)\vert \bar Q\gamma^\mu\gamma_5
 Q\vert 0\rangle=- i f_{P} P^{\mu}\,,\\
&&\langle H(^3S_1,P,\varepsilon^*)\vert \bar Q\gamma^\mu
 Q\vert 0\rangle=- i f_{V} m_V \varepsilon^{*\mu}\,,\\
&&\langle H(^3S_1,P,\varepsilon^*)\vert \bar Q i\sigma^{\mu\nu}
 Q\vert 0\rangle= i f_{V}^{\perp} (\mu)(P^\mu\varepsilon^{*\nu}-P^\nu\varepsilon^{*\mu})\,,\\
&&\langle H(^1P_1,P,\varepsilon^*)\vert \bar Q i\sigma^{\mu\nu}\gamma_5
 Q\vert 0\rangle= -i f_{1A}^{\perp} (\mu)(P^\mu\varepsilon^{*\nu}-P^\nu\varepsilon^{*\mu})\,,\\
&&\langle H(^3P_1,P,\varepsilon^*)\vert \bar Q \gamma^\mu\gamma_5
 Q\vert 0\rangle=- i f_{3A}(\mu) m_{3A}\varepsilon^{*\mu}\,.
\end{eqnarray}

In practical calculations, it is convenient to use the Fourier transformed form of the non-local operator defined in Eq.(\ref{eq:string})
\begin{eqnarray}\label{Pdef}
Q[\Gamma](x) &=& \Big[(\bar Q W_c)(\omega n_+/2)\slash{n}_+\Gamma
(W_c^\dagger Q)(-\omega n_+/2)\Big]_{\rm F.T.}
\nonumber\\
&\equiv& \int \frac{d\omega}{2\pi}\,e^{-i(x-1/2) \omega n_+ P}\, (\bar Q
W_c)(\omega n_+/2)\slash{n}_+\Gamma(W_c^\dagger Q)(-\omega n_+/2), 
\end{eqnarray}
which are invariant under the re-parametrization $n_+\to \alpha n_+$ and $n_-\to \alpha^{-1}n_-$.  We have
\begin{eqnarray}
\langle H(^1S_1,P)\vert Q[\gamma_5](x)\vert 0\rangle&=&- i f_{P} \hat{\phi}_{P}(x)\,,\\
\langle H(^3S_1,P,\varepsilon^*)\vert Q[1](x)\vert 0\rangle&=&- i f_{V} \frac{m_V n_+\varepsilon^*}{n_+P}\hat{\phi}_{V}^{\parallel}(x)\,,\\
\langle H(^3S_1,P,\varepsilon^*)\vert
Q[\gamma^\alpha_\perp](x)\vert 0\rangle&=& -i f_{V}^{\perp}
\varepsilon_\perp^{*\alpha} \hat{\phi}_{V}^{\perp}(x)\,,\\
\langle H(^1P_1,P,\varepsilon^*)\vert Q[\gamma_5](x)\vert 0\rangle&=&- i f_{1A} \frac{m_{1A}{n_+\varepsilon^*}}{n_+P}
\hat{\phi}_{1A}^{\parallel}(x)\,,\\
\langle H(^1P_1,P,\varepsilon^*)\vert
Q[\gamma^\alpha_\perp\gamma_5](x)\vert 0\rangle&=& -i f_{1A}^{\perp}
\varepsilon_\perp^{*\alpha} \hat{\phi}_{1A}^{\perp}(x)\,,\\
\langle H(^3P_0,P)\vert Q[1](x)\vert
0\rangle&=& 
f_{S}  \hat{\phi}_{S}(x)\,,\\
\langle H(^3P_1,P,\varepsilon^*)\vert Q[\gamma_5](x)\vert 0\rangle&=&- i f_{3A} \frac{m_{3A}{n_+\varepsilon^*}}{n_+P}
\hat{\phi}_{3A}^{\parallel}(x)\,,\\
\langle H(^3P_1,P,\varepsilon^*)\vert
Q[\gamma^\alpha_\perp\gamma_5](x)\vert 0\rangle&=& -i f_{3A}^\perp 
\varepsilon_\perp^{*\alpha} \hat{\phi}_{3A}^{\perp}(x)\,,\\
\langle H(^3P_2,P,\varepsilon^*)\vert Q[1](x)\vert
0\rangle&=& 
f_{T} \frac{m_T^2 n_{+\alpha}n_{+\beta}\varepsilon^{*\alpha\beta}}{(n_+P)^2}\hat{\phi}_{T}^{\parallel}(x)\,,\\
\langle H(^3P_2,P,\varepsilon^*)\vert
Q[\gamma^\alpha_\perp](x)\vert 0\rangle&=& f_{T}^{\perp} \frac{m_T n_{+\rho}
\varepsilon^{*\rho\alpha_\perp}}{n_+P} \hat{\phi}_{T}^{\perp}(x)\,.
\end{eqnarray}
Here we suppress the dependence of all quantities on the renormalization scale $\mu$.

\subsection{NRQCD factorization for the LCDAs}

Since quarkonia are non-relativistic bound states of heavy quark and anti quark, all of the LCDAs of quarkonia can be factorized into products of perturbatively calculable distribution parts and non-perturbative NRQCD matrix elements, as what done in \cite{Ma:2006hc,Bell:2008er}. This means that, schematically, at operator level, we have the matching equation
\begin{eqnarray}
 Q[\Gamma](x,\mu)\simeq \sum\limits_{n=0}^{\infty} C^n_\Gamma(x,\mu) O_{\Gamma,n}^{\rm NRQCD}\,,
\end{eqnarray}
where $n$ denotes the order of $v$-expansion,  $C_\Gamma^n(x,\mu)$ is the short-distance coefficient as a distribution over the light-cone fraction $x$, and $O_{\Gamma,n}^{\rm NRQCD}$ is the relevant NRQCD operator which scales ${\cal O}(v^n)$ in the NRQCD power counting . Thus, the LCDAs of quarkonia can be expressed as
\begin{eqnarray}\label{eq:matching}
\langle H\vert Q[\Gamma](x,\mu)\vert 0\rangle \simeq \sum\limits_{n=0}^{\infty} C^n_\Gamma(x,\mu) \langle H\vert O_{\Gamma,n}^{\rm NRQCD}\vert 0\rangle\,.
\end{eqnarray}

At the lowest order of $v$, the matrix elements of the following relevant NRQCD effective operators will be involved in our calculation:
\begin{eqnarray}
\label{eq:3}
\mathcal{O}(^1S_0)&\equiv& \bar\psi_v
\gamma_5 \chi_v\,,\nn\\
\mathcal{O}^\mu(^3S_1)&\equiv&\bar\psi_v
 \gamma_{\top}^{\mu} \chi_v\,,\nn\\
\mathcal{O}^\mu (^1P_1)&\equiv&
\bar\psi_v \left[\left(-\frac{i}{2}\right)
\stackrel{\leftrightarrow}{D}_{\top}^{\mu}\gamma_5\right] \chi_v\,, \nn\\
\mathcal{O}(^3P_0)&\equiv&
\bar\psi_v \left[-\frac{1}{\sqrt{3}}\left(-\frac{i}{2}\right)
\stackrel{\leftrightarrow}{\Slash D}_{\top}\right] \chi_v\,, \\
\nn
\mathcal{O}^{\rho\mu\nu}(^3P_1)&\equiv&\frac{1}{2\sqrt{2}} 
\bar\psi_v \left(-\frac{i}{2}\right)
\stackrel{\leftrightarrow}{D}_{\top}^{\rho}\left[\gamma_\top^\mu,\gamma_{\top}^{\nu}
\right]\gamma_5 \chi_v\,,\\
\mathcal{O}^\mu(^3P_1)&\equiv&\frac{1}{2\sqrt{2}} 
\bar\psi_v \left(-\frac{i}{2}\right)
\left[\stackrel{\leftrightarrow}{\Slash D}_{\top},\gamma_{\top}^{\mu}
\right]\gamma_5 \chi_v\,,\nn\\
\mathcal{O}^{\mu\nu}(^3P_2)&\equiv&
\bar\psi_v \left[\left(-\frac{i}{2}\right)
\stackrel{\leftrightarrow}{D}_{\top}^{(\mu} \gamma_{\top}^{\nu)}\right] \chi_v\nn \,.
\end{eqnarray}
Here we use the four-component notations as in \cite{beneke} for the NRQCD Lagrangian, 
\begin{eqnarray}\label{eq:NRQCDL}
{\cal L}_{\rm NRQCD}^{\rm LO}=\bar\psi_v\left(iv\cdot D-\frac{\left(i D^{\mu}_\top\right)\left(i D_{\top\mu} \right)}{2 m}\right)\psi_v+\bar\chi_v\left(iv\cdot D+\frac{\left(i D^{\mu}_\top\right)\left(i D_{\top\mu} \right)}{2 m}\right)\chi_v\,,
\end{eqnarray}
where $m$ is the pole mass of the heavy quark, $\psi_v$ and $\chi_v$ are the effective fields of the heavy-quark and anti-heavy-quark, respectively, satisfying $\slash v\psi_v=\psi_v$ and $\slash v\chi_v=-\chi_v$. $D^\mu= \partial^\mu-ig_s A^\mu$ is the covariant derivative, and $\stackrel{\leftrightarrow}D^{\mu}=\stackrel{\rightarrow}D^\mu-\stackrel{\leftarrow}D^{\mu}=\stackrel{\rightarrow}\partial^\mu-\stackrel{\leftarrow}\partial^{\mu}-2ig_s A^\mu$. And $a^{(\mu}_\top b^{\nu)}_\top\equiv (a_\top^\mu b_\top^\nu+a_\top^\nu b_\top^\mu)/2-a_\top\cdot b_\top (g^{\mu\nu}-v^\mu v^\nu)/(d-1)$ with $d=4$ means the symmetric 3-D traceless part of rank-2 tensor $a^\mu_\top b_\top^\nu$.

At tree-level,  we have \footnote{Here we have used the spin symmetry of heavy quark system to relate the various matrix elements of S-wave operators and P-wave operators.}
\begin{eqnarray}\label{eq:OP}
\left\{\begin{array}{rcl}
\langle \eta_Q |\mathcal{O}(^{1}S_0) |0 \rangle
&=&  \langle \mathcal{O}(^{1}S_0) \rangle,
\\ 
\langle \Psi/\Upsilon|\mathcal{O}^\mu(^{3}S_1) |0\rangle
&=& \varepsilon^{*\mu} \, \langle \mathcal{O}(^{1}S_0) \rangle,\\ 
\langle h_Q |\mathcal{O}^\mu(^{1}P_1) |0 \rangle
&=&  \varepsilon^{*\mu} \,\langle \mathcal{O}(^{3}P_0) \rangle,
\\ 
\langle \chi_{Q0} |\mathcal{O}(^{3}P_0) |0\rangle
&=&  \langle \mathcal{O}(^{3}P_0) \rangle,\\ 
\langle \chi_{Q1} |\mathcal{O}^{\rho\mu\nu} (^{3}P_1) |0\rangle
&=& \frac{1}{2-d}\left(\varepsilon^{*\mu}\left(g^{\rho\nu}-\frac{P^\rho P^{\nu}}{m_{3A}^2}\right)-\varepsilon^{*\nu}\left(g^{\rho\mu}-\frac{P^\rho P^{\mu}}{m_{3A}^2}\right) \right)\, \langle \mathcal{O}(^{3}P_0) \rangle,
\\ 
\langle \chi_{Q1} |\mathcal{O}^\mu (^{3}P_1) |0\rangle
&=& \varepsilon^{*\mu} \, \langle \mathcal{O}(^{3}P_0) \rangle,
\\
\langle \chi_{Q2} |\mathcal{O}^{\mu\nu}(^{3}P_2) |0\rangle
&=&  \varepsilon^{*\mu\nu} \,
 \langle \mathcal{O}(^{3}P_0) \rangle,
\end{array}\right.
\end{eqnarray}
with
\begin{eqnarray}
\label{metowf}
\langle\mathcal{O}(^{1}S_0)\rangle &=&\sqrt{2 N_c} \sqrt{2M_{\eta_{Q}}}
\,\sqrt{\frac{1}{4\pi}} R_{10}(0)\,,
\\
\langle\mathcal{O}(^{3}P_0)\rangle &=&\sqrt{2 N_c} \sqrt{2M_{\chi_{Q0}}}
\,(-i)\,\sqrt{\frac{3}{4\pi}} R^\prime_{21}(0)\,,
\end{eqnarray}
in the color-singlet model at the leading order of $\alpha_s$ and $v$-expansions.
Here $R_{nl}(r)$ denotes the radial
Schr\"odinger wave function of the quarkonium with radial quantum number $n$ and orbit-angular momentum $l$ , and the prime denotes a derivative with the respect of $r$.

\subsection{Tree-level matching} 

The short distance coefficient $C_\Gamma^n(x,\mu)$ can be extracted, most conveniently, through matching the matrix-elements between the vacuum and state of a colorless pair of free heavy quark and anti quark with non-relativistic relative motion.  In this subsection, we illustrate how to do the matching at tree level. The generalization to the NLO calculation is straightforward. 

We start with the heavy quark and anti-quark pair with the momenta
\begin{eqnarray}
&&p_1^\mu =m v^\mu+q^\mu\,,~~p_2^\mu =m v^\mu+\tilde{q}^\mu\,,~~p_{1,2}^2=m^2\,,
\end{eqnarray}  
where the residual momenta $q$ and $\tilde q$ in the rest frame of heavy quark pair scale like $q^0=\tilde q^0\sim mv^2$, $\vec{q}=-\vec{\tilde q}$ and $\vert \vec{q}\vert=\vert \vec{\tilde q}\vert \sim mv$, where $v<<1$. The total momentum of heavy quark pair
\begin{eqnarray}
P^\mu=p_1^\mu+p_2^\mu=m_H v^\mu\,,~~m_H\equiv \sqrt{P^2}\approx 2 m+{\cal O}(v^2)\,.
\end{eqnarray}
The on-shell spinors of quark and anti-quark can be expanded in $v$ as
\begin{eqnarray}
u(p_1)\approx \left(1+\frac{\slash q}{2 m}\right) u_v(p_1)\,,~~~u_v(p_1)\equiv\frac{1+\slash v}{2} u(p_1)\,,\\\
v(p_2)\approx \left(1-\frac{ \tilde{\slash q}}{2 m}\right) v_v(p_2)\,,~~~v_v(p_2)\equiv\frac{1-\slash v}{2} v(p_2)\,.
\end{eqnarray}

Thus,  at tree-level, we have 
\begin{eqnarray}
&&\langle  Q^a(p_1)\bar Q^b(p_2) \vert Q[\Gamma](x)\vert 0\rangle \nonumber \\
&=&\delta^{ab}\int \frac{d\omega}{2\pi}\,e^{-i (x-1/2) \omega  n_+ P+i \omega  n_+\bar q}\bar u(p_1)\slash n_+\Gamma v(p_2)\nonumber\\
&=&\frac{ \delta^{ab}}{n_+P}\delta\left(x-1/2-\frac{n_+\bar q}{n_+P}\right)\bar u(p_1) \slash n_+\Gamma v(p_2)\nonumber\\
&=& \frac{\delta^{ab} }{n_+P}\left(\left(\delta(x-1/2) -\delta^\prime(x-1/2)\frac{n_+\bar q}{n_+P}\right)\bar u_v(p_1) \slash n_+\Gamma v_v(p_2)\right.\nonumber\\
&&\left.+\delta(x-1/2)\frac{1}{2 m}\bar u_v(p_1) \left\{\bar \slash q,\slash n_+\Gamma\right\}v_v(p_2)+{\cal O}(v^2)\right)\,,
\end{eqnarray}
where we define $\bar q\equiv (q-\tilde q)/2$, and $a,b$ are color indices for the quark and anti-quark. 

For illustration, when $\Gamma=\gamma_5$, we have 
\begin{eqnarray}
&&\bar u_v(p_1)\slash n_+ \gamma_5 v_v(p_2)=n_+v\bar u_v(p_1)\gamma_5 v_v(p_2)\sim {n_+v}\langle Q\bar Q\vert  \mathcal{O}(^1S_0)\vert 0\rangle \,,\\
&& \frac{1}{2 m}\bar u_v(p_1) \left\{\bar \slash q,\slash n_+\gamma_5\right\}v_v(p_2)=\frac{n_{+\mu}}{2m}\bar u_v(p_1)[\bar \slash q,\gamma_\top^\mu]\gamma_5 v_v(p_2)\sim 
\frac{\sqrt{2}}{m}\langle Q\bar Q\vert n_{+\mu}\mathcal{O}^\mu(^3P_1)\vert 0\rangle\,,\nn\\\\
&& \frac{n_+\bar {q}}{n_+P}\bar u_v(p_1)\slash n_+ \gamma_5 v_v(p_2)=\frac{n_+vn_+\bar q}{n_+P}\bar u_v(p_1) \gamma_5 v_v(p_2)\sim \frac{n_+v}{n_+P} \langle Q\bar Q\vert n_{+\mu}\mathcal{O}^\mu(^1P_1)\vert 0\rangle \,.
\end{eqnarray}
Thus, 
\begin{eqnarray}
\langle Q[\gamma_5](x)\rangle &=&\delta(x-1/2)\left(\frac{n_+v}{n_+P}\langle \mathcal{O}(^1S_0)\rangle +\frac{\sqrt{2}n_{+\mu}}{ mn_+P}\langle \mathcal{O}^\mu(^3P_1)\rangle\right)\nn\\
&&-\frac{\delta^\prime(x-1/2)}{n_+P}\frac{n_+vn_{+\mu}}{n_+P}\langle \mathcal{O}^\mu(^1P_1)\rangle +{\cal O}(v^2)\,.
\end{eqnarray} 
With the normalization conditions for the LCDAs set by  (\ref{eq:norm1},\ref{eq:norm2}),  we have
\begin{eqnarray}
\hat\phi_P^{(0)}(x)=\hat\phi_{3A}^{\parallel(0)}(x)=\delta(x-1/2)\,,~~\hat\phi_{1A}^{\parallel(0)}(x)=-\delta^\prime(x-1/2)/2\,,
\end{eqnarray}
and
\begin{eqnarray}
&& f_P^{(0)}=\frac{i}{m_P}\langle {\cal O}(^1S_0)\rangle\,,~~ f_{1A}^{(0)}=i\frac{2}{m_{1A}^2}\langle\mathcal{O}(^3P_0)\rangle\,,~~f_{3A}^{(0)}=i\frac{\sqrt{2}}{mm_{3A}}\langle\mathcal{O}(^3P_0)\rangle\,,
\end{eqnarray}
where the superscript  $(0)$ denotes the quantity at the leading order of $\alpha_s$. Note we have used the fact that $n_+v/n_+P=1/m_H$.

Similarly, one can get
\begin{eqnarray}
&&\hat\phi_V^{\parallel(0)}(x)=\hat\phi_{V}^{\perp(0)}(x)=\hat\phi_{1A}^{\perp(0)}(x)=\delta(x-1/2)\,,\\
&&\hat\phi_{S}^{(0)}(x)=\hat\phi_{3A}^{\perp^{(0)}}(x)=\hat\phi_{T}^{\parallel(0)}(x)=\hat\phi_{T}^{\perp(0)}(x)=-\delta^\prime(x-1/2)/2\,,\end{eqnarray}
and
\begin{eqnarray}
&&f_V^{(0)}=f_V^{\perp(0)}=\frac{i}{m_V}\langle \mathcal{O}(^1S_0)\rangle\,,~~f_{1A}^{\perp(0)}=-\frac{i}{m m_{1A}}\langle\mathcal{O}(^3P_0)\rangle\,,\\
&&f_{S}^{(0)}=-\frac{2}{\sqrt{3}m_{S}^2}\langle\mathcal{O}(^3P_0)\rangle\,,~~ f_{3A}^{\perp(0)}=-\frac{i\sqrt{2}}{m_{3A}^2}\langle\mathcal{O}(^3P_0)\rangle\,,\\
&&f_{T}^{(0)}=-f_{T}^{\perp(0)}=-\frac{2}{m_{T}^2}\langle\mathcal{O}(^3P_0)\rangle\,.
\end{eqnarray}

\section{The calculations of the LCDAs at NLO\label{sect:NLOLCDAs}}

\subsection{Matching procedure by method of threshold expansion}
To extract the short-distance coefficients $C^n_\Gamma(x,\mu)$ at NLO of $\alpha_s$ through the matching equation (\ref{eq:matching}), we have to calculate one-loop corrections to the matrix elements of both $Q[\Gamma](x,\mu)$ and $O_{\Gamma,n}^{\rm NRQCD}$ in general matching procedure as what done in \cite{Ma:2006hc}. 

However, in this work, we will adopt the method of threshold expansion \cite{Beneke:1997zp} to simplify the matching procedure so that we do not need to calculate the one-loop corrections to the matrix elements of effective operators $O_{\Gamma,n}^{\rm NRQCD}$. This is equivalent to what done in \cite{Bell:2008er}. 

In Feynman gauge, at one-loop level, the bare matrix element of $Q[\Gamma](x)$ is written as \footnote{Here we set the momentum of gluon in the loop as $k-\bar q$ as in \cite{Beneke:1997zp}.  And note that $p_{1,2}=P/2\pm \bar q$, the momenta of quark and anti-quark propagators will be $k+P/2$ and $P/2-k$, respectively. }
\begin{eqnarray}\label{eq:bare}
&&\langle Q^a(p_1) \bar Q^b(p_2)\vert Q[\Gamma](x)\vert 0\rangle^{\rm bare}=\delta^{ab}\delta\left(x-\frac{n_+p_1}{n_+P}\right)\frac{\bar u(p_1)\slash n_+\Gamma v(p_2)}{n_+P}\nn\\
&&+\frac{\alpha_s}{4\pi}C_F\delta^{ab}\int[dk]\frac{\bar u(p_1)\gamma^\mu(\slash k+\Slash P/2+m)\slash n_+\Gamma (\slash k-\Slash P/2+m)\gamma_\mu v(p_2)}{n_+P[(k-\bar q)^2][(k+P/2)^2-m^2][(k-P/2)^2-m^2]}\delta\left(x-\frac{1}{2} -\frac{n_+k}{n_+P}\right)\nn\\
&&-\frac{\alpha_s}{4\pi}C_F\delta^{ab}\int [dk]\frac{\bar u(p_1)\slash n_+(\slash k+\Slash P/2+m)\slash n_+\Gamma  v(p_2)\left(\delta\left(x-\frac{1}{2}-\frac{n_+k}{n_+P}\right)-\delta\left(x-\frac{n_+p_1}{n_+P}\right)\right)}{n_+P n_+(k-\bar q)[(k-\bar q)^2][(k+P/2)^2-m^2]}\nn\\
&&-\frac{\alpha_s}{4\pi}C_F\delta^{ab}\int [dk]\frac{\bar u(p_1)\slash n_+\Gamma (\slash k-\Slash P/2+m)\slash n_+ v(p_2)\left(\delta\left(x-\frac{1}{2}-\frac{n_+k}{n_+P}\right)-\delta\left(x-\frac{n_+p_1}{n_+P}\right)\right)}{n_+P n_+(k-\bar q)[(k-\bar q)^2][(k-P/2)^2-m^2]}\,,\nn\\
\end{eqnarray}
where $+i\epsilon$  prescription for the propagators is understood,  $\alpha_s=g_s^2/(4\pi)^2$ is the running strong coupling,
 $C_F=\frac{N_c^2-1}{2N_c}$ with $N_c=3$ is rank-2 Casimir in the fundamental representation of SU(3) group, and
\[[dk]\equiv \frac{(4\pi)^2}{i}\left(\frac{e^{\gamma_E}\mu^2}{4\pi}\right)^\varepsilon\frac{d^dk}{(2\pi)^d}\,,\]
with $d=4-2\varepsilon$ and $\gamma_E=0.5772...$ being the Euler constant. In the following calculations, we will use the dimensional regularization (DR) to regulate both of the ultraviolet and infrared divergences.

Apparently, we have to fix the scheme to treat $\gamma_5$ 
in DR. In the literature, two schemes about $\gamma_5$ in DR are widely-used, one is the naive dimensional regularization (NDR) scheme  \cite{Chanowitz:1979zu}, in which 
 $\{\gamma_5,\gamma^\mu\}=0$, $\{\gamma^\mu,\gamma^\nu\}=2 g^{\mu\nu}$ and $g_\mu^\mu=d$; 
the other is the t'Hooft-Veltman (HV) scheme \cite{'tHooft:1972fi,Breitenlohner:1977hr}, in which $\gamma_5\equiv i\gamma^0\gamma^1\gamma^2\gamma^3$, and $\{\gamma^\mu,\gamma_5\}=0$ for $\mu=0,1,2,3$ but $[\gamma^\mu,\gamma_5]=0$ for $\mu=4,...,d-1$. In this paper, we will compute the NLO corrections to the LCDAs in both the NDR and HV schemes. 

The commonly used method to deal with the spinor bilinear $\bar u(p_1)\cdots v(p_2)$ in NRQCD community, is to transform it into a trace of Dirac matrices ${\rm Tr}[v(p_2)\bar u(p_1)\cdots]$
by replacing $v(p_2)\bar u(p_1)$ with the proper spin-singlet or spin-triplet projectors. In many cases, the $\gamma_5$ involved trace is unavoidable. 
In contrast to the HV scheme, in which such traces involving $\gamma_5$ are defined uniquely and consistently, the NDR scheme for traces involving $\gamma_5$ are generally ill-defined.  Thus, the additional care should be paid in evaluating the odd-number of $\gamma_5$s involved trace. For instance, in \cite{Korner:1991sx} the authors proposed a strategy to treat traces involving an odd number of $\gamma_5$s in the NDR scheme, by which one can easily reproduce the celebrated Adler-Bell-Jakiw 
anomaly, and other $\gamma_5$ involved loop calculations that are consistent with those obtained in the HV scheme. 

However, in this paper, we will not use the trace techniques to calculate the spinor bilinear $\bar u(p_1)\cdots v(p_2)$. In general, we have to deal with a spinor bilinear like
\begin{eqnarray}
\bar u(p_1)\cdots \slash n_+\Gamma\cdots v(p_2)\,,
\end{eqnarray} 
where $\slash n_+\Gamma$ originates from the vertex of  $Q[\Gamma](x)$, and the ellipses denote
complex of Dirac matrices product from the QCD vertex and quark propagators. As we have seen in Sect.\ref{sect:definitions}, $\Gamma=1,\gamma_5,\gamma_\perp^\alpha,\gamma_\perp^\alpha\gamma_5$, and we set $n_{\pm}^\mu$, $v^\mu$, $\gamma^\alpha$ and both the external momenta within 4 dimensions.  Then, no matter in the NDR or HV scheme, $\slash n_{\pm}$ either commutes or anti-commutes with $\Gamma$ from $Q[\Gamma]$. The loop momentum $k$ can be decomposed into 
\begin{eqnarray}
k^\mu=n_+k \frac{n_-^\mu}{2}+n_-k\frac{n_+^\mu}{2}+k_\perp^\mu\,,
\end{eqnarray}
in which $k_\perp^\mu$ can run over the extra dimensions $\mu=4,...,d-1$. Therefore, (\ref{eq:bare}) can be simplified to


\begin{eqnarray}
&&\langle Q^a(p_1) \bar Q^b(p_2)\vert Q[\Gamma](x)\vert 0\rangle^{\rm bare}=\delta^{ab}\delta\left(x-\frac{n_+p_1}{n_+P}\right)\frac{\bar u(p_1)\slash n_+\Gamma v(p_2)}{n_+P}\nn\\
&&+\frac{\alpha_s}{4\pi}C_F\delta^{ab}\int[dk]\frac{\bar u(p_1)\gamma^\mu(\slash k+\Slash P/2+m)\slash n_+\Gamma (\slash k-\Slash P/2+m)\gamma_\mu v(p_2)}{n_+P[(k-\bar q)^2][(k+P/2)^2-m^2][(k-P/2)^2-m^2]}\delta\left(x-\frac{1}{2} -\frac{n_+k}{n_+P}\right)\nn\\
&&-\frac{\alpha_s}{4\pi}C_F\delta^{ab}\int [dk]\frac{2n_+( k+P/2)\left(\delta\left(x-\frac{1}{2}-\frac{n_+k}{n_+P}\right)-\delta\left(x-\frac{n_+p_1}{n_+P}\right)\right)}{n_+P n_+(k-\bar q)[(k-\bar q)^2][(k+P/2)^2-m^2]}\bar u(p_1)\slash n_+\Gamma  v(p_2)\nn\\
&&-\frac{\alpha_s}{4\pi}C_F\delta^{ab}\int [dk]\frac{2n_+( k-P/2)\left(\delta\left(x-\frac{1}{2}-\frac{n_+k}{n_+P}\right)-\delta\left(x-\frac{n_+p_1}{n_+P}\right)\right)}{n_+P n_+(k-\bar q)[(k-\bar q)^2][(k-P/2)^2-m^2]}\bar u(p_1)\slash n_+\Gamma  v(p_2)\,,
\end{eqnarray}
where implicitly
\[[dk]\equiv \frac{(4\pi)^2}{i}\left(\frac{e^{\gamma_E}\mu^2}{4\pi}\right)^\varepsilon\frac{dn_+kd^{d-2}k_\perp dn_-k}{2(2\pi)^d}\,.\]

We will expand the loop integrals in small parameter $v\sim \vert \vec{ q}\vert/m$ by the threshold expansion technique developed in \cite{Beneke:1997zp}. The most important momentum regions are hard region (where loop momentum $k^\mu\sim m$),  soft region (where $k^\mu\sim m v$), potential region (where $k^\mu\sim m(v^2, \vec{v})$), ultra-soft region (where $k^\mu\sim m v^2$). The contributions from the low-energy regions, i.e. (ultra)-soft and potential regions, are reproduced by the one-loop corrections to the matrix elements of effective operators in matching equation (\ref{eq:matching}). Thus, to get the NLO part of the short distance coefficient $C_{\Gamma,n}(x,\mu)$, we only need to calculate the contributions from the hard region. 

After the tedious expansions of integrands in hard region, we get various complicated spinor bilinears with complicated spin-structures. At first, we try to use only identities $\{\gamma^\mu,\gamma^\nu\}=2g^{\mu\nu}$, $\{\slash n_{\pm},\Gamma\}=0$ or $[\slash n_{\pm},\Gamma]=0$ and on-shell conditions for the external spinors as much as possible, for which identities hold in both NDR and HV schemes, to simplify the spin-structures. And in the end, it turns out that the only possible $\gamma_5$-dependent structures are\footnote{Here, we would like to emphasize that, even though we use spinor decomposition technique instead of trace technique, the NDR scheme is still algebraically inconsistent in contrast to the HV scheme. One can possiblly obtain the different results with different manipulations or strategies for spin-structure simplifications. The strategy of calculations in the NDR scheme in this paper, which are conventional in the literatures, such as in \cite{Beneke:2004rc,Beneke:2005gs}, is to try using  identities, such as $\{\gamma^\mu,\gamma^\nu\}=2 g^{\mu\nu}$ and the on-shell conditions of external momenta as much as possible, and identify the scheme-dependent spin-structures (here the spin-structures listed in (\ref{eq:gamma})) for final treatment that relies on the anti-commuting properties of $\gamma_5$ in the NDR scheme.  }
\begin{eqnarray}\label{eq:gamma}
\gamma^\rho \slash n_+\Gamma\gamma_\rho~{\rm and }~\gamma^\rho \gamma^\sigma\slash n_+\Gamma\gamma_\sigma\gamma_\rho\,.
\end{eqnarray}
We define $\gamma^\rho \slash n_+\Gamma \gamma_\rho \equiv c_{\slash n_+\Gamma} \slash n_+\Gamma$ so that
\begin{eqnarray}
c_{\slash n_+\Gamma}=\left\{
\begin{array}{ll}
2-d\,, &~~~\Gamma=\mathbf{1}\,,\\
d-2\,, &~~~\Gamma=\gamma_5\,,\\
d-4\,, &~~~\Gamma=\gamma^\alpha_\perp\,,\\
4-d\,, &~~~\Gamma=\gamma^\alpha_\perp\gamma_5\,,
\end{array}
\right.
\end{eqnarray}
in the NDR scheme and 
\begin{eqnarray}
c_{\slash n_+\Gamma}=\left\{
\begin{array}{ll}
2-d\,, &~~~\Gamma=\mathbf{1}\,,\\
6-d\,, &~~~\Gamma=\gamma_5\,,\\
d-4\,, &~~~\Gamma=\gamma^\alpha_\perp\,,\\
d-4\,, &~~~\Gamma=\gamma^\alpha_\perp\gamma_5\,,
\end{array}
\right.
\end{eqnarray}
in the HV scheme. 

Thus, the hard part of the bare matrix element up to ${\cal O}(v)$ is
\begin{eqnarray}
&&\langle Q^a(p_1) \bar Q^b(p_2)\vert Q[\Gamma](x)\vert 0\rangle^{\rm bare}_{\rm hard}\nonumber\\
&=&\delta^{ab}\frac{\bar u_v(p_1)\slash n_+\Gamma v_v(p_2)}{n_+P}\left\{\delta\left(x-\frac{1}{2}\right)\right.\nonumber\\
&&+\frac{\alpha_s}{4\pi}C_F\int[dk]\frac{\delta\left(x-\frac{1}{2}-\frac{n_+k}{n_+P}\right )\left(-4m^2+c_{\slash n_+\Gamma}^2k_\perp^2/(d-2)-c_{\slash n_+\Gamma}\left(n_+k/n_+v\right)^2\right)}{[k^2+i\epsilon][k^2+P\cdot k+i\epsilon][k^2-P\cdot k+i\epsilon]}\nn\\
&&\left.-\frac{\alpha_s}{4\pi}C_F\int [dk]\frac{\delta\left(x-\frac{1}{2}-\frac{n_+k}{n_+P}\right)-\delta\left(x-\frac{1}{2}\right)}{ [n_+k][k^2+i\epsilon]}\left(\frac{2n_+( k+P/2)}{ [k^2+P\cdot k+i\epsilon]}+\frac{2n_+( k-P/2)}{ [k^2-P\cdot k+i\epsilon]}\right)\right\}\nn\\
&&+\frac{\delta^{ab}}{2m}\frac{\bar u_v(p_1)\left\{\bar\slash  q,\slash n_+\Gamma\right\} v_v(p_2)}{n_+P}\left\{\delta\left(x-\frac{1}{2}\right)\right.\nonumber\\
&&+\frac{\alpha_s}{4\pi}C_F\int[dk]\frac{\delta\left(x-\frac{1}{2}-\frac{n_+k}{n_+P}\right )\left((c_{\slash n_+\Gamma}^2-8m^2/k^2)k_\perp^2/(d-2)+c_{\slash n_+\Gamma}\left(n_+k/n_+v\right)^2\right)}{[k^2+i\epsilon][k^2+P\cdot k+i\epsilon][k^2-P\cdot k+i\epsilon]}\nonumber\\
&&\left.-\frac{\alpha_s}{4\pi}C_F\int [dk]\frac{\delta\left(x-\frac{1}{2}-\frac{n_+k}{n_+P}\right)-\delta\left(x-\frac{1}{2}\right)}{ [n_+k][k^2+i\epsilon]}\left(\frac{2n_+( k+P/2)}{ [k^2+P\cdot k+i\epsilon]}+\frac{2n_+( k-P/2)}{ [k^2-P\cdot k+i\epsilon]}\right)\right\}\nonumber\\
&&+\delta^{ab}\frac{n_+\bar q}{n_+P}\frac{\bar u_v(p_1)\slash n_+\Gamma v_v(p_2)}{n_+P}\left\{-\delta^\prime\left(x-\frac{1}{2}\right)\right.\nonumber\\
&&+\frac{\alpha_s}{4\pi}C_F\int[dk]\frac{n_+P\left(n_-k-\frac{n_+k}{(n_+v)^2}\right)\delta\left(x-\frac{1}{2}-\frac{n_+k}{n_+P}\right )}{[k^2+i\epsilon]^2[k^2+P\cdot k+i\epsilon][k^2-P\cdot k+i\epsilon]}\nonumber\\
&&~~~~~~\times \left(-4m^2+c_{\slash n_+\Gamma}^2\frac{k_\perp^2}{d-2}-c_{\slash n_+\Gamma}\left(\frac{n_+k}{n_+v}\right)^2\right)\nonumber\\
&&-\frac{\alpha_s}{4\pi}C_F\int[dk]\frac{\delta\left(x-\frac{1}{2}-\frac{n_+k}{n_+P}\right )4c_{\slash n_+\Gamma}m n_+k/n_+v}{[k^2+i\epsilon][k^2+P\cdot k+i\epsilon][k^2-P\cdot k+i\epsilon]}\nonumber\\
&&+\frac{\alpha_s}{4\pi}C_F\int[dk]\frac{8mc_{\slash n_+\Gamma} (n_+k/n_+v)k_\perp^2/(d-2) \delta\left(x-\frac{1}{2}-\frac{n_+k}{n_+P}\right )}{[k^2+i\epsilon]^2[k^2+P\cdot k+i\epsilon][k^2-P\cdot k+i\epsilon]}\nonumber\\
&&-\frac{\alpha_s}{4\pi}C_F\int [dk]\frac{\delta^\prime\left(x-\frac{1}{2}\right)}{ [n_+k][k^2+i\epsilon]}\left(\frac{2n_+( k+P/2)}{ [k^2+P\cdot k+i\epsilon]}+\frac{2n_+( k+P/2)}{ [k^2+P\cdot k+i\epsilon]}\right)\nn\\
&&-\frac{\alpha_s}{4\pi}C_F\int [dk]\frac{\delta\left(x-\frac{1}{2}-\frac{n_+k}{n_+P}\right)-\delta\left(x-\frac{1}{2}\right)}{ [n_+k][k^2+i\epsilon]}\left(\frac{2n_+( k+P/2)}{ [k^2+P\cdot k+i\epsilon]}+\frac{2n_+( k+P/2)}{ [k^2+P\cdot k+i\epsilon]}\right)\nonumber\\
&&~~~~~~~~~~~~\left.\times\left(\frac{n_+P\left(n_-k-\frac{n_+k}{(n_+v)^2}\right)}{ [k^2+i\epsilon]}+\frac{n_+P}{ [n_+k]}\right)\right\}\nn\\
&&+\frac{\delta^{ab}}{2m}\frac{\bar u_v(p_1)\left[\bar\slash  q,\slash n_+\Gamma\right] v_v(p_2)}{n_+P}\frac{\alpha_s}{4\pi}C_F\int[dk]\frac{\delta\left(x-\frac{1}{2}-\frac{n_+k}{n_+P}\right )}{[k^2+i\epsilon][k^2+P\cdot k+i\epsilon][k^2-P\cdot k+i\epsilon]}\nonumber\\
&&\times\left\{4m^2 \frac{n_+k}{n_+v}\left(-2+c_{\slash n_+\Gamma}-2c_{\slash n_+\Gamma} \frac{k_\perp^2}{(d-2)k^2}\right)\right\}+{\cal O}(v^2)\,.
\end{eqnarray}

The hard part of the renormalized matrix-element is
\begin{eqnarray}
&&\langle Q^a(p_1)\bar Q^b(p_2)\vert Q[\Gamma](x)\vert 0\rangle^{\rm ren}_{\rm hard}=Z^{\rm os}_2 \int_0^1 dy Z_{\slash n_+ \Gamma}(x,y)\langle Q^a(p_1)\bar Q^b(p_2)\vert Q[\Gamma](y)\vert 0\rangle^{\rm bare}_{\rm hard}\,,
\end{eqnarray}
where the on-shell renormalization constant for the heavy quark is
\begin{eqnarray}
Z_{2}^{\rm
os}&=&1-\frac{\alpha_s}{4\pi}C_F\left(\frac{3}{\varepsilon}+3\ln\frac{\mu^2}{m^2}+4\right)\,,
\end{eqnarray}
and the renormalization kernels for the operator $Q[\Gamma](x)$ in the $\overline{\rm MS}$ scheme are
\begin{eqnarray}
Z_{\slash n_+\gamma_5}(x,y)&=&Z_{\slash n_+}(x,y)\,=\,\delta(x-y)-\frac{\alpha_s}{4\pi}C_F \frac{2}{\varepsilon} V_0(x,y)\,,\\
Z_{\slash n_+\gamma_\perp^\alpha\gamma_5}(x,y)&=&Z_{\slash n_+ \gamma_\perp^\alpha}(x,y)\,=\,\delta(x-y)-\frac{\alpha_s}{4\pi}C_F \frac{2}{\varepsilon} V_\perp(x,y)\,,
\end{eqnarray}
with the Brodsky-Lepage kernel being
\begin{eqnarray}
V_0(x,y)&=&\left[\frac{1- x}{1-
y}\left(1+\frac{1}{x-y}\right)\theta(x-y)+\frac{x}{y}\left(1+\frac{1}{y-x}\right)
\theta(y-x)\right]_+\,,\\
V_\perp(x,y)&=&V_0(x,y)-\left[\frac{1- x}{1-
y}\theta(x-y)+\frac{x}{y} \theta(y-x)\right]\,.
\end{eqnarray}

Therefore, schematically, the final matching equation up to ${\cal O}(v)$ goes to
\begin{eqnarray}\label{eq:matching1}
&&\langle Q^a(p_1)\bar Q^b(p_2)\vert Q[\Gamma](x)\vert 0\rangle^{\rm ren}_{\rm hard}\nonumber\\
&=& \sum\limits_{n=0,1}C_{\Gamma,n}(x,\mu)\langle Q^a(p_1)\bar Q^b(p_2)\vert \mathcal{O}^{\rm NRQCD}_n\vert 0\rangle_{\rm tr} +{\cal O}(v^2)\,.
\end{eqnarray}

 Before we close the description of our matching procedure, one last thing we have to mention, is that in general covariant gauge, we should get additional contributions to (\ref{eq:bare}). However, since we are calculating the on-shell matrix elements of gauge invariant operators, such additional contributions should vanish in the end. And we check that, by our strategy to simplify the spin-structures, no matter whether we are in the NDR scheme or HV scheme, such additional terms in general covariant gauge do vanish, as they should. This guarantees the gauge invariance of our results.

\subsection{Final results for LCDAs of quarkonia}

Giving the concrete $\Gamma$ in (\ref{eq:matching1}), we can simplify the spin-structures further, and decompose them into the matrix elements of the effective operators in (\ref{eq:3}), as we did in the previous section. By use of the loop integrals given in Appendix \ref{app:loop}, we obtain the short-distance coefficients $C_{\Gamma,n}(x,\mu)$.  Imposing the normalization conditions given in (\ref{eq:norm1}) and (\ref{eq:norm2}), we reach the final results for the LCDAs at the NLO of $\alpha_s$ and leading order of $v$. 

The three LCDAs for the S-wave quarkonia are 
\begin{eqnarray}
\hat\phi_P(x;\mu)&=&\delta(x-1/2)+\frac{\alpha_s(\mu)}{4\pi}C_F\left\{4\left[\left(\ln\frac{\mu^2}{m^2(1-2
x)^2}-1\right)\left(1+\frac{1}{1/2-x}\right)x\theta(1-2
x)\right]_{+}\right.\nonumber\\
&&\left.+\left[\frac{16
x\bar x}{(1- 2x)^2}\theta(1-2x)\right]_{++}+\Delta\left[16x\theta(1-2x)\right]_++(x\leftrightarrow \bar x)\right\}\,,\\
\hat\phi_V^\parallel(x;\mu)&=&\delta(x-1/2)+\frac{\alpha_s(\mu)}{4\pi}C_F\left\{4\left[\left(\ln\frac{\mu^2}{m^2(1-2
x)^2}-1\right)\left(1+\frac{1}{1/2-x}\right)x\theta(1-2
x)\right]_{+}\right.\nonumber\\
&&\left.+\left[\frac{16
x\bar x}{(1- 2x)^2}\theta(1-2x)\right]_{++}-\left[8x\theta(1-2x)\right]_++(x\leftrightarrow \bar x)\right\}\,,\\
\hat\phi_V^\perp(x;\mu)&=&\delta(x-1/2)+\frac{\alpha_s(\mu)}{4\pi}C_F\left\{\left[\left(\ln\frac{\mu^2}{m^2(1-2
x)^2}-1\right)\frac{8x}{1-2x}\theta(1-2
x)\right]_{+}\right.\nonumber\\
&&\left.+\left[\frac{16
x\bar x}{(1- 2x)^2}\theta(1-2x)\right]_{++}+(x\leftrightarrow \bar x)\right\}\,,
\end{eqnarray}
and the corresponding decay constants are
\begin{eqnarray}
f_P&=&\left\{1+\frac{\alpha_s(\mu)}{4\pi}C_F\left(-6+4\Delta\right)\right\}\frac{i}{m_P}\langle {\cal O}(^1S_0)\rangle\,,\\
f_V&=&\left\{1+\frac{\alpha_s(\mu)}{4\pi}C_F\left(-8\right)\right\}\frac{i}{m_V}\langle {\cal O}(^1S_0)\rangle\,,\\
f_V^\perp&=&\left\{1+\frac{\alpha_s(\mu)}{4\pi}C_F\left(-\ln\frac{\mu^2}{m^2}-8\right)\right\}\frac{i}{m_V}\langle {\cal O}(^1S_0)\rangle\,.
\end{eqnarray}
Here, $\Delta=0$ for the NDR scheme, and $\Delta=1$ for the HV scheme.

Similarly, the seven LCDAs for the P-wave quarkonia are
\begin{eqnarray}
\hat\phi_{1A}^\parallel(x;\mu)&=&-\delta^{\prime}(x-1/2)/2\nonumber\\
&&+\frac{\alpha_s}{4\pi}C_F \left\{-\left[\left(\ln\frac{\mu^2}{m^2(1-2x)^2}-3\right)\frac{4x(5-8x+4x^2) \theta(1-2
x)}{(1-2x)^2}\right]_{++}\right.\nn\\
&&-\left[\frac{8x(7-4x) \theta(1-2
x)}{(1-2x)^2}\right]_{++}-\Delta\left[16x \theta(1-2
x)\right]_{++}\nn\\
&&\left.-\left[\frac{8 x\theta(1-2
x)}{(1-2x)^{3}}\right]_{+++}-(x\leftrightarrow \bar x)\right\}\,,\\
\hat\phi_{1A}^\perp(x;\mu)&=&\hat\phi_{V}^\perp(x;\mu)-\frac{\alpha_s(\mu)}{4\pi}C_F\left[\frac{8 x \bar x\theta(2
x-1)}{(1-2x)^{2}}+\frac{8 x \bar x\theta(1-2
x)}{(1-2x)^{2}}\right]_{++}\,,\\
\hat\phi_{S}(x;\mu)&=&-\delta^{\prime}(x-1/2)/2\nn\\
&&+\frac{\alpha_s(\mu)}{4\pi}C_F \left\{-\left[\left(\ln\frac{\mu^2}{m^2(1-2x)^2}-1\right)\frac{4x(5-8x+4x^2) \theta(1-2
x)}{(1-2x)^2}\right]_{++}\right.\nn\\
&&\left.-\left[\frac{8x(7-8x) \theta(1-2
x)}{(1-2x)^2}\right]_{++}-\left[\frac{8 x\theta(1-2
x)}{(1-2x)^{3}}\right]_{+++}-(x\leftrightarrow \bar x)\right\}\,,\\
\hat\phi_{3A}^\parallel(x;\mu)&=&\hat\phi_{V}^\parallel(x;\mu)-\frac{\alpha_s(\mu)}{4\pi}C_F\left[\frac{8 x \bar x\theta(2
x-1)}{(1-2x)^{2}}+\frac{8 x \bar x\theta(1-2
x)}{(1-2x)^{2}}\right]_{++}\nn\\
&&+\Delta\frac{\alpha_s(\mu)}{4\pi}C_F\left[16 x\theta(1-2x)+16\bar x\theta(2x-1)\right]_+\,,\\
\hat\phi_{3A}^\perp(x;\mu)&=&-\delta^{\prime}(x-1/2)/2\nn\\
&&+\frac{\alpha_s(\mu)}{4\pi}C_F \left\{-\left[\left(\ln\frac{\mu^2}{m^2(1-2x)^2}\right)\frac{16x\bar x\theta(1-2
x)}{(1-2x)^2}\right]_{++}-\left[\frac{16x\theta(1-2
x)}{(1-2x)^2}\right]_{++}\right.\nn\\
&&\left.-\left[\frac{8 x\theta(1-2
x)}{(1-2x)^{3}}\right]_{+++}-(x\leftrightarrow \bar x)\right\}\,,\\
\hat\phi_{T}^\parallel(x;\mu)&=&-\delta^{\prime}(x-1/2)/2\nn\\
&&+\frac{\alpha_s(\mu)}{4\pi}C_F \left\{-\left[\left(\ln\frac{\mu^2}{m^2(1-2x)^2}-4\right)\frac{4x(5-8x+4x^2) \theta(1-2
x)}{(1-2x)^2}\right]_{++}\right.\nn\\
&&\left.-\left[\frac{4x(17-10x) \theta(1-2
x)}{(1-2x)^2}\right]_{++}-\left[\frac{8 x\theta(1-2
x)}{(1-2x)^{3}}\right]_{+++}-(x\leftrightarrow \bar x)\right\}\,,\\
\hat\phi_{T}^\perp(x;\mu)&=&-\delta^{\prime}(x-1/2)/2\nn\\
&&+\frac{\alpha_s(\mu)}{4\pi}C_F \left\{-\left[\left(\ln\frac{\mu^2}{m^2(1-2x)^2}+2\right)\frac{16x\bar x\theta(1-2
x)}{(1-2x)^2}\right]_{++}\right.\nn\\
&&\left.+\left[\frac{32x\theta(1-2
x)}{1-2x}\right]_{++}-\left[\frac{8 x\theta(1-2
x)}{(1-2x)^{3}}\right]_{+++}-(x\leftrightarrow \bar x)\right\}\,,
\end{eqnarray}
and the decay constants are
\begin{eqnarray}
f_{1A}&=&\left\{1-\frac{\alpha_s(\mu)}{4\pi}C_F\left(\frac{8}{3}\ln\frac{\mu^2}{m^2}+\frac{76}{9}-\frac{4}{3}\Delta\right)\right\}\frac{2i}{m_{1A}^2}\langle {\cal O}(^3P_0)\rangle\,,\\
f_{1A}^\perp&=&\left\{1+\frac{\alpha_s(\mu)}{4\pi}C_F\left(-\ln\frac{\mu^2}{m^2}-4\right)\right\}\frac{2i}{m_{1A}^2}\langle {\cal O}(^3P_0)\rangle\,,\\
f_{S}&=&\left\{1-\frac{\alpha_s}{4\pi}C_F\left(\frac{8}{3}\ln\frac{\mu^2}{m^2}-\frac{2}{9}\right)\right\}\frac{-2}{\sqrt{3}m_{S}^2}\langle {\cal O}(^3P_0)\rangle\,,\\
f_{3A}&=&\left\{1+\frac{\alpha_s(\mu)}{4\pi}C_F\left(-4+4\Delta \right)\right\}\frac{\sqrt{2}i}{mm_{3A}}\langle {\cal O}(^3P_0)\rangle\,,\\
f_{3A}^\perp&=&\left\{1-\frac{\alpha_s(\mu)}{4\pi}C_F\left(3\ln\frac{\mu^2}{m^2}+6\right)\right\}\frac{-\sqrt{2}i}{m_{3A}^2}\langle {\cal O}(^3P_0)\rangle\,,\\
f_{T}&=&\left\{1-\frac{\alpha_s(\mu)}{4\pi}C_F\left(\frac{8}{3}\ln\frac{\mu^2}{m^2}+\frac{88}{9}\right)\right\}\frac{-2}{m_{T}^2}\langle {\cal O}(^3P_0)\rangle\,,\\
f_{T}^\perp&=&\left\{1-\frac{\alpha_s(\mu)}{4\pi}C_F\left(3\ln\frac{\mu^2}{m^2}+10\right)\right\}\frac{2}{m_{T}^2}\langle {\cal O}(^3P_0)\rangle\,.
\end{eqnarray}

In the above expressions, the $+++$, $++$ and $+$-functions are defined as
\begin{eqnarray}
&&\int_0^1 dx [f(x)]_{+++}g(x)=\int_0^1 dx f(x)(g(x)-g(1/2)-g^\prime (1/2)(x-1/2)\nonumber\\
&&~~~~~~~~~~~~~~~~~~~~~~~~~~~~~~~~~-\frac{g^{\prime\prime} (1/2)}{2}(x-1/2)^2),\\
&&\int_0^1 dx [f(x)]_{++}g(x)=\int_0^1 dx f(x)(g(x)-g(1/2)-g^\prime (1/2)(x-1/2)),\\
&&\int_0^1 dx [f(x)]_{+}g(x)=\int_0^1 dx f(x)(g(x)-g(1/2))\,.
\end{eqnarray}

One can check that our results for $\hat \phi_M(x;\mu)$ preserve the normalizations in (\ref{eq:norm1},\ref{eq:norm2}), and $f_M\phi_M(x;\mu)$ satisfy the ERBL equations
\begin{eqnarray}
\mu^2\frac{d}{d\mu^2}\left(f_M
\hat\phi_M(x)\right)&=&\frac{\alpha_s(\mu^2)}{2\pi}C_F\int_0^1 dy
V_M(x,y)\left(f_M \hat\phi_M(y)\right)\,.
\end{eqnarray}

For the decay constants which can be defined by the local QCD currents, such as $f_P,f_V,f_V^\perp,f_{1A}^\perp$ and $f_{3A}$, we find that our results at NLO of $\alpha_s$  and in the NDR scheme agree with those in literature \cite{Kniehl:2006qw}. The decay constants, such as $f_{1A}^\perp,f_S,f_{3A}^\perp,f_{T}$, and $f_{T}^\perp$, are actually the first Gegenbauer moments of the corresponding LCDAs, which satisfy the renormalization group equations that they should obey \cite{Lepage:1979zb,Efremov:1979qk}, 
\begin{eqnarray}
\frac{d}{d\ln\mu^2}(f_S,f_{3A},f_T)&=&-\frac{\alpha_s}{4\pi}C_F\left(\frac{8}{3}\right)(f_S,f_{3A},f_T)\,,\\
\frac{d}{d\ln\mu^2}(f_{1A}^\perp,f_T^\perp)&=&-\frac{\alpha_s}{4\pi}C_F(3)(f_{1A}^\perp,f_T^\perp)\,.
\end{eqnarray}

We also compare our results for the LCDAs of S-wave quarkonia with those in \cite{Ma:2006hc,Bell:2008er}. In \cite{Ma:2006hc}, the authors give all three leading twist LCDAs for S-wave quarkonia, but we find that their results do not lead to correct decay constants at NLO of $\alpha_s$ after integration over the light fraction either in the NDR scheme or in the HV scheme. In \cite{Bell:2008er}, only $f_P\hat{\phi}_P(x)$ is calculated, and we find that our result in the NDR scheme agrees with theirs.

\subsection{Some related quantities}
In the practical applications of the leading twist LCDAs, since the lowest order hard-kernels $T_H(x)$ for many hard exclusive processes are in form of $1/x$ or $1/\bar x$, the inverse moments of the LCDAs are crucial for final amplitudes. 

We define 
\begin{eqnarray}
R_\Gamma\equiv\frac{f_\Gamma}{f_\Gamma^{(0)}}\,,~~\Gamma=P,V_\parallel,V_\perp,1A_\parallel,1A_\perp,S,3A_\parallel,3A_\perp,T_\parallel,T_\perp\,.
\end{eqnarray}
We have
\begin{eqnarray}
&&R_P\int_0^1 dx \hat\phi_P(x,\mu)\frac{1}{x}=2+\frac{\alpha_s}{4\pi}C_F\left((6-4\ln 2)\ln\frac{\mu^2}{m^2}+4(1+4\Delta)\ln 2-\frac{2\pi^2}{3}\right)\,,\\
&&R_V^\parallel\int_0^1 dx \hat\phi_V^\parallel(x,\mu)\frac{1}{x}=2+\frac{\alpha_s}{4\pi}C_F\left((6-4\ln 2)\ln\frac{\mu^2}{m^2}-4\ln 2-\frac{2\pi^2}{3}\right)\,,
\\&&R_{3A}^\parallel\int_0^1 dx \hat\phi_{3A}^\parallel(x,\mu)\frac{1}{x}=2+\frac{\alpha_s}{4\pi}C_F\left((6-4\ln 2)\ln\frac{\mu^2}{m^2}-4(1-4\Delta)\ln 2+4-\frac{2\pi^2}{3}\right)\,,\nn\\
\\&&R_V^\perp\int_0^1 dx \hat\phi_V^\perp(x,\mu)\frac{1}{x}=2+\frac{\alpha_s}{4\pi}C_F\left((6-8\ln 2)\ln\frac{\mu^2}{m^2}+8\ln 2-\frac{4\pi^2}{3}\right)\,,
\\
&&R_{1A}^\perp\int_0^1 dx \hat\phi_{1A}^\perp(x,\mu)\frac{1}{x}=2+\frac{\alpha_s}{4\pi}C_F\left((6-8\ln 2)\ln\frac{\mu^2}{m^2}+8\ln 2+4-\frac{4\pi^2}{3}\right)\,,\\
&&R_{T}^\parallel\int_0^1 dx \hat\phi_{T}^\parallel(x,\mu)\frac{1}{x}=-2+\frac{\alpha_s}{4\pi}C_F\left(\left(-2+4\ln 2\right)\ln\frac{\mu^2}{m^2}+4\ln 2+\frac{2\pi^2}{3}\right)\,,
\\
&&R_S\int_0^1 dx \hat\phi_{S}(x,\mu)\frac{1}{x}=-2+\frac{\alpha_s}{4\pi}C_F\left(\left(-2+4\ln 2\right)\ln\frac{\mu^2}{m^2}-20\ln 2-12+\frac{2\pi^2}{3}\right)\,,
\\&&R_{1A}^\parallel\int_0^1 dx \hat\phi_{1A}^\parallel(x,\mu)\frac{1}{x}=-2+\frac{\alpha_s}{4\pi}C_F\left(\left(-2+4\ln 2\right)\ln\frac{\mu^2}{m^2}+(4\ln 2-4)(1+4\Delta)+\frac{2\pi^2}{3}\right)\,,\nn\\\\
&&R_{3A}^\perp\int_0^1 dx \hat\phi_{3A}^\perp(x,\mu)\frac{1}{x}=-2+\frac{\alpha_s}{4\pi}C_F\left(2\ln\frac{\mu^2}{m^2}+8\ln 2-8\right)\,,
\\
&&R_{T}^\perp\int_0^1 dx \hat\phi_{T}^\perp(x,\mu)\frac{1}{x}=-2+\frac{\alpha_s}{4\pi}C_F\left(2\ln\frac{\mu^2}{m^2}+24\ln 2-8\right)\,,\end{eqnarray}
with $\Delta=0$ for the NDR scheme and $\Delta=1$ for the HV scheme.

\section{Applications\label{sect:applications}}
In this section, we will apply our results for the LCDAs of quarkonia to calculate the hard exclusive processes $\gamma^*\to
\eta_Q \gamma,\chi_{QJ}\gamma$, $Z\to \eta_Q\gamma,\chi_{QJ}\gamma, J/\psi(\Upsilon) \gamma, h_Q\gamma$ and   $h\to J/\psi\gamma$ within the collinear factorization\footnote{ In \cite{Jia:2008ep}, the authors have considered the $\gamma^*\to \eta_b  \gamma$ and $h\to \Upsilon \gamma$ in the collinear factorization at the leading logarithm level, and used ERBL equations to resum the large logarithms. The applications in this paper are a kind of extension of the work in \cite{Jia:2008ep} at the NLO of $\alpha_s$, but we shall not consider the resummation here.}. We also compare our results with the asymptotic behavior of the corresponding predictions  in the NRQCD factorization. These comparisons can be regarded as a non-trivial test of our results.

\subsection{$\gamma^*\to \eta_Q  \gamma,\chi_{QJ}\gamma$ in the collinear factorization}

For the hard exclusive process $\gamma^*(Q,\varepsilon_{\gamma^*})\to H(p) \gamma(p^\prime,\varepsilon_\gamma)$ with the momenta in the light-cone coordinates 
\begin{eqnarray}
p^{\prime\mu}=n_-p^\prime \frac{n_+^\mu}{2}\,,~~~p^\mu=n_+p\frac{n_-^\mu}{2}+n_-p\frac{n_+^\mu}{2}\,,
\end{eqnarray}
and the polarization vectors $\epsilon_{\gamma^*}$ and $\epsilon_{\gamma}$ for the virtual and real photon, respectively, 
when $m_H^2/Q^2<<1$ ($Q^2=(p+p^\prime)^2$), we expect the light-cone factorization formula for the transition amplitude
\begin{eqnarray}
&& i{\cal M}(\gamma^*(Q,\varepsilon_{\gamma^*})\to H(p) \gamma(p^\prime,\varepsilon_\gamma))\nonumber\\
&=&-ie^2 e_Q^2\varepsilon_{\gamma^*\mu}\varepsilon_{\gamma\nu}^*\int_0^1 dx\left (\frac{i}{2}\epsilon^{\mu\nu}_\perp  T_H^P(x;Q^2,\mu)\langle H(p)\vert Q[\gamma_5](x;\mu)\vert 0\rangle\right.\nonumber\\
&&~~~~~~~\left.+\frac{g^{\mu\nu}_\perp }{2}T_H^V(x;Q^2,\mu)\langle H(p)\vert Q[1](x;\mu)\vert 0\rangle\right)+{\cal O}(m_H^2/Q^2)\,.
\end{eqnarray}
Here $e$ is the elementary electric charge,  $e_Q$ the fractional electric charge of the quark $Q$ inside of meson $H$,  $\epsilon_\perp^{\mu\nu}\equiv \epsilon^{\mu\nu\rho\sigma}n_{-\rho}n_{+\sigma}/2$ where $\epsilon^{\mu\nu\rho\sigma}$ is the Levi-Cevita tensor with $\epsilon_{0123}=+1$, $T_H^{P,V}(x;Q^2,\mu)$ are the perturbatively calculable hard-kernels, the matrix-elements of $Q[\gamma_5](x;\mu)$ and $Q[1](x;\mu)$ are eventually the appropriate leading-twist LCDAs of meson $H$. 

In \cite{Braaten:1982yp}, the hard-kernels have been obtained at the NLO of $\alpha_s$ which are
\begin{eqnarray}
T_H^{P}(x;Q^2,\mu)&=&\frac{1}{\bar x}\left\{1+\frac{\alpha_s}{4\pi}C_F \left[-(3+2 \ln \bar
x)\ln\frac{\mu^2}{-Q^2-i\epsilon}+\ln^2\bar x-\frac{\bar x \ln\bar
x}{x} -9\right]\right\}\nonumber\\&&+\left(x\leftrightarrow \bar x\right)\,,\label{eq:Braaten1}\\
T_H^{V}(x;Q^2,\mu)&=&\frac{1}{\bar x} \left\{1+\frac{\alpha_s}{4\pi}C_F \left[-(3+2 \ln \bar
x)\ln\frac{\mu^2}{-Q^2-i\epsilon}+\ln^2\bar x-3\frac{\bar x \ln\bar
x}{x} -9\right]\right\}\nonumber\\&&-\left(x\leftrightarrow \bar x\right)
\label{eq:Braaten2}\,.
\end{eqnarray}

Actually we did a recalculation of the hard-kernels $T_H^{P,V}$ by using evanescent operator technique proposed in \cite{Herrlich:1994kh}, and obtained the same results as in  \cite{Braaten:1982yp} if we adopt the NDR scheme to treat $\gamma_5$.  
For the problem we consider here, the evanescent operator is
\begin{eqnarray}
Q_E^{\mu\nu}(x;\mu)&\equiv&Q[[\gamma_\perp^\mu,\gamma_\perp^\nu]/2](x;\mu)-\frac{i\epsilon_\perp^{\mu\nu}}{2}Q[\gamma_5](x;\mu)\,,
\end{eqnarray}
which tree-level matrix-element vanishes in 4-dimension, but can contribute a term proportional to $d-4$ in $d$-dimensional loop-calculation in general. If the one-loop coefficient of the tree-level matrix-element $Q_E^{\mu\nu}$ contains a pole in term of $1/\varepsilon$, an additional finite renormalization is required to make sure the matrix-element of $Q_E^{\mu\nu}$ at one-loop level vanishes in the end \cite{Herrlich:1994kh}. In the NDR scheme,  tree-level matrix-element of $Q_E^{\mu\nu}$ does not vanish in $d$-dimension, thus we are required to do the corresponding finite renormalization. However, in the HV scheme,  tree-level matrix-element of $Q_E^{\mu\nu}$ does vanish even in $d$-dimension, so that we do not need to do the additional finite renormalization. This leaves us a great convenience to get the hard-kernels in the HV scheme, even before we get those in the NDR scheme. Thus, in the HV scheme, the hard-kernels read as
\begin{eqnarray}
T_H^{P}(x;Q^2,\mu)&=&\frac{1}{\bar x}\left\{1+\frac{\alpha_s}{4\pi}C_F \left[-(3+2 \ln \bar
x)\ln\frac{\mu^2}{-Q^2-i\epsilon}+\ln^2\bar x+2\frac{\bar x \ln\bar
x}{x} +5\ln x-9\right]\right\}\nonumber\\&&+\left(x\leftrightarrow \bar x\right)\,,\\
T_H^{V}(x;Q^2,\mu)&=&\frac{1}{\bar x} \left\{1+\frac{\alpha_s}{4\pi}C_F \left[-(3+2 \ln \bar
x)\ln\frac{\mu^2}{-Q^2-i\epsilon}+\ln^2\bar x+2\frac{\bar x \ln\bar
x}{x} +5\ln x -9\right]\right\}\nonumber\\&&-\left(x\leftrightarrow \bar x\right)
\,.
\end{eqnarray}
Note that $T_H^V$ in the HV scheme is actually identical to $T_H^V$ in (\ref{eq:Braaten2}), but $T_H^P$ in the HV scheme is different from $T_H^P$ in (\ref{eq:Braaten1}).

Straightforwardly, we apply the LCDAs of quarkonia obtained in the previous section, we have the NLO amplitudes
\begin{eqnarray}
&& i{\cal M}(\gamma^*(Q,\varepsilon_{\gamma^*})\to \eta_Q(p) \gamma(p^\prime,\varepsilon_\gamma))=\frac{i}{2}e^2 e_Q^2\epsilon^{\mu\nu}_\perp \varepsilon_{\gamma^*\mu}\varepsilon_{\gamma\nu}^* f_{\eta_Q}\int_0^1 dx T_H^P(x;Q^2,\mu)\hat{\phi}_P(x;\mu)\nonumber\\
&=&-2e^2 e_Q^2\epsilon^{\mu\nu}_\perp \varepsilon_{\gamma^*\mu}\varepsilon_{\gamma\nu}^* \frac{\langle {\cal O}(^1S_0)\rangle}{m_{\eta_Q}}\left\{1+\frac{\alpha_s}{4\pi}C_F\left[(3-2\ln 2)L+\ln^2 2+3\ln 2-9-\frac{\pi^2}{3}\right]\right\}\,,\nonumber\\ \\ 
&& i{\cal M}(\gamma^*(Q,\varepsilon_{\gamma^*})\to \chi_{Q1}(p,\varepsilon) \gamma(p^\prime,\varepsilon_\gamma))=\frac{i}{2}e^2 e_Q^2\epsilon^{\mu\nu}_\perp \varepsilon_{\gamma^*\mu}\varepsilon_{\gamma\nu}^* f_{3A}\int_0^1 dx  T_H^P(x;Q^2,\mu)\hat{\phi}_{3A}^\parallel(x;\mu)\nonumber\\
&=&-2\sqrt{2}e^2 e_Q^2\epsilon^{\mu\nu}_\perp \varepsilon_{\gamma^*\mu}\varepsilon_{\gamma\nu}^* \frac{\langle {\cal O}(^3P_0)\rangle}{m_{\chi_{Q1}}m}\left\{1+\frac{\alpha_s}{4\pi}C_F\left[(3-2\ln 2)L+\ln^2 2-\ln 2-7-\frac{\pi^2}{3}\right]\right\}\,,\nonumber\\\\
&& i{\cal M}(\gamma^*(Q,\varepsilon_{\gamma^*})\to \chi_{Q0}(p) \gamma(p^\prime,\varepsilon_\gamma))=-i e^2 e_Q^2 \varepsilon_{\gamma^*}\cdot\varepsilon_{\gamma}^* \frac{f_{S}}{2}\int_0^1 dx  T_H^V(x;Q^2,\mu)\hat{\phi}_S(x;\mu)\nonumber\\
&=&i4e^2 e_Q^2\epsilon^{\mu\nu}_\perp \varepsilon_{\gamma^*\mu}\varepsilon_{\gamma\nu}^* \frac{\langle {\cal O}(^3P_0)\rangle}{\sqrt{3}m_{\chi_{Q0}}^2}\left\{1+\frac{\alpha_s}{4\pi}C_F\left[(1-2\ln 2)L+\ln^2 2+9\ln 2-\frac{\pi^2}{3}\right]\right\}\,,\\
&& i{\cal M}(\gamma^*(Q,\varepsilon_{\gamma^*})\to \chi_{Q2}(p) \gamma(p^\prime,\varepsilon_\gamma))=-i e^2 e_Q^2 \varepsilon_{\gamma^*}\cdot\varepsilon_{\gamma}^* \frac{f_{T}}{\sqrt{6}}\int_0^1 dx  T_H^V(x;Q^2,\mu)\hat{\phi}_T^\parallel(x;\mu)\nonumber\\
&=&-i8e^2 e_Q^2\epsilon^{\mu\nu}_\perp \varepsilon_{\gamma^*\mu}\varepsilon_{\gamma\nu}^* \frac{\langle {\cal O}(^3P_0)\rangle}{\sqrt{6}m_{\chi_{Q2}}^2}\left\{1+\frac{\alpha_s}{4\pi}C_F\left[(1-2\ln 2)L+\ln^2 2-3\ln 2-6-\frac{\pi^2}{3}\right]\right\}\,,\nonumber\\
\end{eqnarray}
with $L\equiv \ln\frac{-Q^2-i\epsilon}{m^2}$. One can check
that, although both of the hard-kernels  and LCDAs are dependent on the $\gamma_5$ schemes in loop calculations,  the amplitudes of $\gamma^*\to \eta_Q\gamma$ and $\gamma^*\to \chi_{Q1}\gamma$ are independent of the schemes of $\gamma_5$ as they should be. 

By squaring the amplitudes, one can easily reproduce the asymptotic behavior of the ratios between the NLO  and tree-level cross-sections of $e^+e^-\to \eta_c \gamma, \chi_{cJ}\gamma (J=0,1,2)$ in \cite{Sang:2009jc}. The authors adopted the trace technique proposed in  \cite{Korner:1991sx}. Since only one $\gamma_5$ is involved in the trace, their results are essentially consistent with the results obtained in the HV scheme. 

\subsection{$Z\to \eta_Q  \gamma,\chi_{QJ}\gamma,J/\psi (\Upsilon)\gamma,h_Q\gamma$ in the collinear factorization}

The $Z$ boson interacts with quark-anti-quark pair through the tree-level weak interaction as
\begin{eqnarray}
i{\cal L}_{ZQ\bar Q}=i\frac{g}{4\cos\theta_W}\bar Q\gamma^\mu (g_V-g_A\gamma_5)Q Z_\mu\,,
\end{eqnarray}
where $g$ is the weak coupling in $SU(2)_L\times U(1)_Y$ electro-weak gauge theory, $\theta_W$ the Weinberg angle, $g_V=1-8\sin^2\theta_W/3$ and $g_A=1$ for the up-type quark, and $g_V=-1+4\sin^2\theta_W/3$ and $g_A=-1$ for the down-type quark. 

Thus, through the vectorial interaction, $Z$ can decay to $\eta_Q  \gamma,\chi_{QJ}\gamma$ as $\gamma^*$, the corresponding decay amplitudes in the light-cone framework are just similar to $\gamma^*\to H\gamma$ by replacing the prefactor $e^2 e_Q^2$ with $ g g_V e e_Q/(4\cos\theta_W)$,  $\varepsilon_{\gamma^*}$ with the polarization vector of $Z$ boson $\varepsilon_Z$, and $Q^2$ with $m_Z^2$. Through the axial-vectorial interaction, $Z$ can decay to $J/\psi(\Upsilon) \gamma, h_{Q}\gamma$ as well. The corresponding factorization formula can be reached similarly, i.e.
\begin{eqnarray}
&& i{\cal M}(Z(Q,\varepsilon_{Z})\to H(p) \gamma(p^\prime,\varepsilon_\gamma))\nonumber\\
&=&i \frac{g_A e e_Q}{4\cos\theta_W}\varepsilon_{Z\mu}\varepsilon_{\gamma\nu}^*\int_0^1 dx\left (\frac{i}{2}\epsilon^{\mu\nu}_\perp  \tilde{T}_H^V(x;m_Z^2,\mu)\langle H(p)\vert Q[1](x;\mu)\vert 0\rangle\right.\nonumber\\
&&~~~~~~~\left.+\frac{g^{\mu\nu}_\perp}{2} \tilde{T}_H^A(x;m_Z^2,\mu)\langle H(p)\vert Q[\gamma_5](x;\mu)\vert 0\rangle\right)+{\cal O}(m_H^2/m_Z^2)\,,
\end{eqnarray}
where $H=J/\psi(\Upsilon)$ or $h_Q$, and $\tilde{T}^{V,A}$ are the hard-kernels. 

In the NDR scheme, $\gamma_5$ is anti-commuting with all $\gamma^\mu$. Thus, the hard-kernels $\tilde{T}_H^{V,A}$ can be obtained very quickly, by identifying 
\begin{eqnarray}
\tilde{T}_H^V(x;m_Z^2,\mu)=T_H^P(x;m_Z^2,\mu)\,,~~~\tilde{T}_H^A(x;m_Z^2,\mu)=T_H^V(x;m_Z^2,\mu)\,,\label{eq:thva}
\end{eqnarray}
 where the NLO expressions of $T_H^{P,V}$ in the NDR scheme are given in (\ref{eq:Braaten1},\ref{eq:Braaten2}).

In the HV scheme, the extractions of $\tilde{T}_H^{V,A}$ by adopting the evanescent operator technique \cite{Herrlich:1994kh}, are much more involved than extractions of $T_H^{P,V}$ for $\gamma^*\to H\gamma$, since $\gamma_5$ appears explicitly in the interaction vertex. However, it is straightforward but tedious. In the end, we get the hard-kernels in the HV scheme which read as
\begin{eqnarray}
\tilde{T}_H^{V}(x;m_Z^2,\mu)&=&\frac{1}{\bar x}\left\{1+\frac{\alpha_s}{4\pi}C_F \left[-(3+2 \ln \bar
x)\ln\frac{\mu^2}{-Q^2-i\epsilon}+\ln^2\bar x-\frac{\bar x \ln\bar
x}{x} -9\right]\right\}\nonumber\\&&+\left(x\leftrightarrow \bar x\right)\,,\label{eq:tdv}\\
\tilde{T}_H^{A}(x;m_Z^2,\mu)&=&\frac{1}{\bar x} \left\{1+\frac{\alpha_s}{4\pi}C_F \left[-(3+2 \ln \bar
x)\ln\frac{\mu^2}{-m_Z^2-i\epsilon}+\ln^2\bar x+5\frac{\bar x \ln\bar
x}{x} -9\right]\right\}\nonumber\\&&-\left(x\leftrightarrow \bar x\right)
\label{eq:tda}\,.
\end{eqnarray}
Note that $\tilde{T}_H^V$ in the HV scheme is actually identical to $\tilde{T}_H^V$ in (\ref{eq:thva}), but $\tilde{T}_H^A$ in the HV scheme is different from $\tilde{T}_H^A$ in (\ref{eq:thva}).

Straightforwardly, we have the NLO amplitudes
\begin{eqnarray}
&& i{\cal M}(Z(Q,\varepsilon_{Z})\to  J/\psi (\Upsilon)(p,\varepsilon) \gamma(p^\prime,\varepsilon_\gamma))\nonumber\\
&=&-i\frac{g_A e e_Q}{4\cos\theta_W}\epsilon^{\mu\nu}_\perp \varepsilon_{Z\mu}\varepsilon_{\gamma\nu}^* f_{J/\psi(\Upsilon)}\int_0^1 dx  \tilde{T}_H^V(x;Q^2,\mu)\hat{\phi}_V(x;\mu)\nonumber\\
&=&-i\frac{g_A e e_Q}{2\cos\theta_W}\epsilon^{\mu\nu}_\perp \varepsilon_{Z\mu}\varepsilon_{\gamma\nu}^*  \frac{\langle {\cal O}(^1S_0)\rangle}{m_{{J/\psi(\Upsilon)}}}\left\{1+\frac{\alpha_s}{4\pi}C_F\left[(3-2\ln 2)L+\ln^2 2-\ln 2-9-\frac{\pi^2}{3}\right]\right\}\,,\nonumber\\\\
&& i{\cal M}(Z(Q,\varepsilon_{Z})\to h_{Q}(p,\varepsilon) \gamma(p^\prime,\varepsilon_\gamma))=\frac{g_A e e_Q}{8\cos\theta_W}\varepsilon_{Z}\cdot\varepsilon_{\gamma}^* f_{1A}\int_0^1 dx  \tilde{T}_H^A(x;Q^2,\mu)\hat{\phi}_{1A}^\parallel(x;\mu)\nonumber\\
&=&-i\frac{g_A e e_Q}{2\cos\theta_W}\varepsilon_{Z}\cdot\varepsilon_{\gamma}^*  \frac{\langle {\cal O}(^3P_0)\rangle}{m_{h_{Q}}m}\left\{1+\frac{\alpha_s}{4\pi}C_F\left[(1-2\ln 2)L+\ln^2 2-3\ln 2-4-\frac{\pi^2}{3}\right]\right\}\,,\nonumber\\
\end{eqnarray}
with $L\equiv \ln\frac{-m_Z^2-i\epsilon}{m^2}$. One can also check that the amplitude for $Z\to h_Q\gamma$ is independent of the scheme to treat $\gamma_5$.

By squaring the amplitudes, one should easily reproduce the asymptotic behavior of the ratios between the NLO  and tree-level cross-sections of $e^+e^-\to J/\psi \gamma, h_{c}\gamma$ at $Z^0$-pole.  In \cite{Chen:2013itc,Wu}, Chen {\it et al} give the asymptotic ratios between the NLO and LO cross section are
\begin{eqnarray}
&&r[^3S_1]\equiv \frac{\sigma^{\rm NLO}(e^+e^-\to Z^0\to J/\psi \gamma)}{\sigma^{\rm LO}(e^+e^-\to Z^0\to J/\psi \gamma)}\nn\\
&=&\frac{\alpha_s}{2\pi}C_F\left[(3-2\ln 2)\ln\frac{m_Z^2}{m_c^2}+\ln^2 2-\ln 2-5-\frac{\pi^2}{3}\right]\,,\\
&&r[^1P_1]\equiv \frac{\sigma^{\rm NLO}(e^+e^-\to Z^0\to h_c \gamma)}{\sigma^{\rm LO}(e^+e^-\to Z^0\to h_c \gamma)}\nn\\
&=&\frac{\alpha_s}{2\pi}C_F\left[(1-2\ln 2)\ln\frac{m_Z^2}{m_c^2}+\ln^2 2-3\ln 2-4-\frac{\pi^2}{3}\right]\,.
\end{eqnarray}
Their results agree with ours for $^1P_1$ case, but differ from ours for $^3S_1$ case, by a constant term (-4) at ${\cal O}(\alpha_s)$.  We cannot figure out the source of this discrepancy.

\subsection{$h\to J/\psi \gamma$ in the collinear factorization}

The higgs boson $h$ in the Standard Model interacts with quark-anti-quark pair through the Yukawa interaction
\begin{eqnarray}
i{\cal L}_{hQ\bar Q}=i\frac{y_Q}{\sqrt{2}}\bar QQ h\,.
\end{eqnarray}
Here $y_Q\equiv -\sqrt{2} \overline{m}/v$ is the Yukawa coupling where $v=246$ GeV is the vacuum expectation value of the Higgs field, and $\overline{m}$ is the current mass of quark $Q$ in $\overline{\rm MS}$ scheme.  The corresponding factorization formula for $h\to J/\psi \gamma$ is
\begin{eqnarray}
&& i{\cal M}(h(Q)\to J/\psi(p,\varepsilon_\psi) \gamma(p^\prime,\varepsilon_\gamma))\nonumber\\
&=&-i \frac{m_c e e_c}{2v}\varepsilon_{\gamma\nu}^*\int_0^1 dx T_H(x;m_h^2,\mu)\langle J/\psi(p,\varepsilon_\psi)\vert Q[\gamma_\perp^\nu](x;\mu)\vert 0\rangle\,,
\end{eqnarray}
where $\varepsilon_\psi$ is the polarization vector of $J/\psi$, and the hard-kernel $T_H$ can be calculated perturbatively. The NLO hard-kernel is 
\begin{eqnarray}
T_H(x;m_h^2,\mu)&=&\frac{1}{\bar x}\left\{1+\frac{\alpha_s}{4\pi}C_F\left[-3 \ln\frac{\mu^2}{m_c^2}-2\frac{\ln\bar x}{x}\ln\frac{\mu^2}{-m_h^2-i\epsilon}+\frac{\ln^2\bar x}{x}-7\right]\right\}\nonumber\\
&&+(x\leftrightarrow \bar x)\,,
\end{eqnarray}
with the mass of higgs in the Standard Model $m_h\simeq 125$ GeV. 

Straightforwardly, we have the NLO amplitudes
\begin{eqnarray}\label{eq:higgs}
&& i{\cal M}(h(Q)\to  J/\psi (p,\varepsilon_\psi) \gamma(p^\prime,\varepsilon_\gamma))
\\
&=&-i\frac{m_c e e_c}{2v} \varepsilon_{\psi}^*\cdot\varepsilon_{\gamma}^*  \frac{\langle {\cal O}(^1S_0)\rangle}{m_{{J/\psi}}}\left\{1-\frac{\alpha_s}{4\pi}C_F\left[4\ln2 \ln\frac{-m_h^2-i\epsilon}{m_c^2}-2\ln^2 2-4\ln 2+7+\frac{2\pi^2}{3}\right]\right\}\,,\nonumber
\end{eqnarray}
where $m_c$ is the pole mass of charm quark.

Thirty years ago, Shifman {\it et al} \cite{Shifman:1980dk} had calculated $h\to J/\psi \gamma$ to NLO of $\alpha_s$ in color singlet model which is equivalent to the NRQCD calculation. The NLO prediction for $h\to J/\Psi
\gamma$, that we quote from equation (21) in \cite{Shifman:1980dk}, is
written as
\begin{eqnarray}
&&i{\cal M}(h\to J/\Psi \gamma)=i{\cal M}_{\rm tr}(h\to J/\Psi \gamma)\left[1-\frac{\alpha_s(m_h^2)C_F}{2\pi}a(\kappa)\right]\,,
\end{eqnarray}
where
\begin{eqnarray}
&&a(\kappa)=4-\frac{\pi^2}{12(1-\kappa)}-\frac{F(1-2
\kappa)}{2(1-\kappa)}+\frac{\kappa-1}{1-2\kappa}+\frac{2\kappa(\kappa-2)}{(1-\kappa)^2}\left[\phi(\kappa)+F(1)-F(-1)\right]
\nonumber\\
&&+\left(4+\frac{4}{\kappa}+\frac{8}{1-\kappa}\right)\sqrt{\frac{\kappa}{1-\kappa}}
\arctan\sqrt{\frac{\kappa}{1-\kappa}}+\left(\frac{4}{1-\kappa}+2+\frac{\kappa}{(2\kappa-1)^2}\right)\ln(2-2\kappa)\,,\nonumber\\
\end{eqnarray}
with $\kappa=m_h^2/(4 m_c^2)+i\epsilon$ and 
\begin{eqnarray}
\phi(x)&=&\int_0^1 \frac{dy}{y-1/(2 x)}\ln\frac{1-4 y(1-y) x}{2
y(1-x)}\,,~~F(x)=\int_0^x dy\,\frac{\ln(1+y)}{y}=-{\rm Li}_2(-x)\,.
\end{eqnarray}

The asymptotic behavior of
$a(\kappa)$ at $\kappa\to \infty$ is
\begin{eqnarray}
a(\kappa)&=&\frac{1}{2}\left[4\ln 2\ln\frac{-m_h^2-i\epsilon}{m_c^2}-2\ln^2 2-4
\ln 2+\frac{2\pi^2}{3}+7\right]+{\cal O}(m_c^2/m_h^2)\,,
\end{eqnarray}
which coincides with (\ref{eq:higgs}).

\section{Summary\label{sect:summary}}
In this paper, we calculate ten leading twist LCDAs for the S-wave and P-wave quarkonia to the NLO of $\alpha_s$ and leading order of $v$, in both NDR and HV schemes.  We demonstrate that applications of these LCDAs in some single quarkonium exclusive processes can lead to correct asymptotic behavior of relevant NRQCD results. This confirms again the conclusion in \cite{Jia:2010fw} that there is a tight connection between the collinear factorization method and NRQCD factorization method for a certain class of quarkonium exclusive productions. And also as in \cite{Jia:2008ep}, together with the ERBL equation, the collinear factorization method can be used to resum the large logarithms in NRQCD calculations. However, as discussed in \cite{Jia:2008ep,Bodwin:2013ys}, the so-called "endpoint logarithms" in helicity-flipped exclusive processes,  lead to the breakdown of the collinear factorization. Such "endpoint logarithms" seem to be process-dependent, and how to resum them remains unknown. 

\section*{Acknowledgement}
The authors thank Prof. Yu Jia for enormous inspiring discussions on many issues related to this work, and also thank G. Chen, X.-G.~Wu, Z. Sun, X.-C. Zheng and J.-M. Shen (authors of \cite{Chen:2013itc}) for providing us the asymptotic expansion of their results on $e^+e^-\to J/\psi\gamma, h_c\gamma$ at $Z^0$ pole. This work is partially supported by the National Natural Science Foundation of China under Grants No. 11275263 and No. 10935012.

\vspace{.5cm}

{\it Note added:} After this work was finished,  we were noticed by the authors of  a series of paper \cite{Ma:2013yla,Ma:2014eja}, that they calculated the S-wave and P-wave heavy quarkonium fragmentation functions (FFs) from a heavy quark pair, of which the FFs from a color-singlet heavy quark pair are related to the LCDAs we calculated in this paper.  We are very grateful to Y.Q. Ma, J.W. Qiu and H. Zhang (the authors of \cite{Ma:2013yla,Ma:2014eja}) for enormous communications and efforts on cross-checking.  After correcting some typos and mistakes in original manuscripts, we get completely consistent results in the NDR scheme.

Here we express our special appreciation to Prof. M. Neubert and Prof. G. T. Bodwin for pointing out a few mistakes appearing in the published version in JHEP, in which the explicit NLO expression for $\phi_V^\parallel(x;\mu)$ is missing, and the logarithmic term in NLO hard-kernel for $h\to J/\psi\gamma$ is wrong. We correct them in this updated arXiv version.

\appendix
\section{Some useful integrals \label{app:loop}} 
Here we list some loop integrals which are useful for the NLO computation of LCDAs for quarkonia in Sect.\ref{sect:NLOLCDAs}. 

 \begin{eqnarray}
&& \int
 [dk]\frac{f(n_+k)}{[k^2+i\epsilon]^n[k^2+P\cdot k+i\epsilon][k^2-P\cdot k+i\epsilon]}\nonumber\\
 &=&\left(\frac{-1}{m^2}\right)^n\left(\frac{4\pi\tilde
 \mu^2}{m^2}\right)^\varepsilon \frac{\Gamma(n+\varepsilon)}{\Gamma(n+1)}\int_0^1 dy
f((y-1/2) n_+P)\left(\frac{(2 \bar y)^n \theta(2
y-1)}{(1-2y)^{2n+2\varepsilon}}+\frac{(2 y)^n \theta(1-2
y)}{(1-2y)^{2n+2\varepsilon}}\right)\,,\nn\\\\
&& \int
 [dk]\frac{f(n_+k)k_\perp^\mu k_\perp^\nu}{[k^2+i\epsilon][k^2+P\cdot k+i\epsilon][k^2-P\cdot k+i\epsilon]}\nonumber\\
 &=&\left(\frac{4\pi\tilde
 \mu^2}{m^2}\right)^\varepsilon \Gamma(\varepsilon)\frac{g_\perp^{\mu\nu}}{2}\int_0^1 dy
f((y-1/2) n_+P)\left(\frac{2 \bar y \theta(2
y-1)}{(1-2y)^{2\varepsilon}}+\frac{2 y \theta(1-2
y)}{(1-2y)^{2\varepsilon}}\right)\,,\\
&& \int
 [dk]\frac{f(n_+k) k_\perp^\mu  k_\perp^\nu}{[k^2+i\epsilon]^2[k^2+P\cdot k+i\epsilon][k^2-P\cdot k+i\epsilon]}\nonumber\\
 &=&-\frac{1}{m^2}\left(\frac{4\pi\tilde
 \mu^2}{m^2}\right)^\varepsilon \Gamma(1+\varepsilon)\frac{g_\perp^{\mu\nu}}{2}\int_0^1 dy
f((y-1/2) n_+P)\left(\frac{ 2\bar y^2  \theta(2
y-1)}{(1-2y)^{2+2\varepsilon}}+\frac{2y^2 \theta(1-2
y)}{(1-2y)^{2+2\varepsilon}}\right)\,,\nn\\\\
&& \int
 [dk]\frac{f(n_+k)(n_-k-n_+k/(n_+v)^2)n_+\bar q}{[k^2+i\epsilon]^2[k^2+P\cdot k+i\epsilon][k^2-P\cdot k+i\epsilon]}\nonumber\\
 &=&\frac{n_+\bar q}{m^2n_+P}\left(\frac{4\pi\tilde
 \mu^2}{m^2}\right)^\varepsilon \Gamma(1+\varepsilon)\int_0^1 dy
f((y-1/2) n_+P)
\nn\\
&&~~~~~~~~~~~~~~~~~~~\times\left(\frac{4 \bar y(1+2\varepsilon \bar y) \theta(2
y-1)}{(1-2y)^{3+2\varepsilon}}+\frac{4 y(1+2\varepsilon y) \theta(1-2
y)}{(1-2y)^{3+2\varepsilon}}\right)\,,
\\
&& \int
 [dk]\frac{f(n_+k)  k_\perp^\mu  k_\perp^\nu(n_-k-n_+k/(n_+v)^2)n_+\bar q}{[k^2+i\epsilon]^2[k^2+P\cdot k+i\epsilon][k^2-P\cdot k+i\epsilon]}\nonumber\\
 &=&\frac{n_+\bar q}{n_+P}\left(\frac{4\pi\tilde
 \mu^2}{m^2}\right)^\varepsilon \Gamma(\varepsilon)g_\perp^{\mu\nu}\int_0^1 dy
f((y-1/2) n_+P)\nn\\
&&~~~~~~~~~~~~~~~~~\times\left(\frac{ 2\bar y (1-2 y-2 \varepsilon \bar y) \theta(2
y-1)}{(1-2y)^{1+2\varepsilon}}-\frac{2y (1-2 y+2 \varepsilon y) \theta(1-2
y)}{(1-2y)^{1+2\varepsilon}}\right)\,,\\
 &&\int [dk] \frac{\delta(x-1/2-n_+k/n_+P)-\delta(x-1/2)}{[n_+k][k^2+i\epsilon]}\left(\frac{n_+(k+P/2)}{[k^2+P\cdot k+i\epsilon]}+\frac{n_+(k-P/2)}{[k^2-P\cdot k+i\epsilon]}\right)\nn\\
&=&
-\left(\frac{4\pi\tilde
\mu^2}{m^2}\right)^\varepsilon\Gamma(\varepsilon)\left[
\frac{4x \theta(1-2
x)}{(1-2x)^{1+2\varepsilon}}+
\frac{4\bar x \theta(1-2
x)}{(2x-1)^{1+2\varepsilon}}\right]_+\,,\\
 &&n_+\bar q\int [dk] \frac{\delta(x-1/2-n_+k/n_+P)-\delta(x-1/2)}{[n_+k]^2[k^2+i\epsilon]}\left(\frac{n_+(k+P/2)}{[k^2+P\cdot k+i\epsilon]}+\frac{n_+(k-P/2)}{[k^2-P\cdot k+i\epsilon]}\right)\nn\\
&=&
\frac{n_+\bar q}{n_+P}\left(\frac{4\pi\tilde
 \mu^2}{m^2}\right)^\varepsilon\Gamma(\varepsilon)\left[\frac{4x \theta(1-2
x)}{(1-2x)^{2+2\varepsilon}}-\frac{4\bar x \theta(1-2
x)}{(1-2x)^{2+2\varepsilon}}\right]_+\,,\\
&&n_+\bar q\int [dk] \frac{\delta(x-1/2-n_+k/n_+P)-\delta(x-1/2)}{[n_+k][k^2+i\epsilon]^2}\left(n_-k-\frac{n_+k}{(n_+v)^2}\right)\nn\\
&&~~~~~~~~~~~~~~~~~\times \left(\frac{n_+(k+P/2)}{[k^2+P\cdot k+i\epsilon]}+\frac{n_+(k-P/2)}{[k^2-P\cdot k+i\epsilon]}\right)\nn\\
 &=&\frac{n_+\bar q}{n_+P}\left(\frac{4\pi\tilde
 \mu^2}{m^2}\right)^\varepsilon\Gamma(\varepsilon)\left[\frac{8x (1-2 x+4\varepsilon x)\theta(1-2
x)}{(1-2x)^{2+2\varepsilon}}-\frac{8\bar x (1-2 \bar x+4\varepsilon \bar x)\theta(2
x-1)}{(1-2x)^{2+2\varepsilon}}\right]_+\,.
 \end{eqnarray}


\end{document}